\documentclass[fleqn,usenatbib]{mnras}

\usepackage{newtxtext,newtxmath}
\newcommand{\cmark}{$\surd$}%
\usepackage[T1]{fontenc}

\DeclareRobustCommand{\VAN}[3]{#2}
\let\VANthebibliography\thebibliography
\def\thebibliography{\DeclareRobustCommand{\VAN}[3]{##3}\VANthebibliography}

\usepackage{graphicx}	% Including figure files
\usepackage{amsmath}	% Advanced maths commands
\usepackage{booktabs}
\usepackage{tabularx}
\usepackage{marvosym}

\title[Feedback effects on the cosmic SFH in \textsc{Simba}]{The Effects of Stellar and AGN Feedback on the Cosmic Star Formation History in the \textsc{Simba} Simulations
}

\author[L. Scharr\'e et al.]{
Lucie Scharr\'e$^{1,2}$\thanks{E-mail: lucie.scharre@epfl.ch},
Daniele Sorini$^{3, 2}$,
Romeel Dav\'e$^{2,4,5}$
\\
$^{1}$Institute of Physics, Laboratory for Galaxy Evolution, EPFL, Observatoire de Sauverny, Chemin Pegasi 51, 1290 Versoix, Switzerland \\
$^{2}$Institute for Astronomy, University of Edinburgh, Royal Observatory, Edinburgh, EH9 3HJ, United Kingdom \\
$^{3}$Institute for Computational Cosmology, Durham University, South Park Road, DH1 3LE, United Kingdom\\
$^{4}$University of the Western Cape, Bellville, Cape Town 7535, South Africa\\
$^{5}$Stellenbosch Institute for Advanced Studies, Mosterdrift, Stellenbosch 7600, South Africa\\
}

\date{Accepted XXX. Received YYY; in original form ZZZ}

\pubyear{2024}

\begin{document}
\label{firstpage}
\pagerange{\pageref{firstpage}--\pageref{lastpage}}
\maketitle

% Abstract of the paper
\begin{abstract}
Using several variants of the cosmological \textsc{Simba} simulations, we investigate the impact of different feedback prescriptions on the cosmic star formation history. Adopting a global-to-local approach, we link signatures seen in global observables, such as the star formation rate density (SFRD) and the galaxy stellar mass function (GSMF), to feedback effects in individual galaxies. We find a consistent picture: stellar feedback mainly suppresses star formation below halo masses of $M_{\rm H} = 10^{12} \rm \, M_{\odot}$ and before $z = 2$, whereas AGN feedback quenches the more massive systems after $z = 2$. Among \textsc{Simba}'s AGN feedback modes, AGN jets are the dominant quenching mechanism and set the shape of the SFRD and the GSMF at late times. AGN-powered winds only suppress the star formation rate in intermediate-mass galaxies ($M_{\rm \star} = 10^{9.5 - 10} \rm \, M_{\odot}$), without affecting the overall stellar mass-assembly significantly. At late times, the AGN X-ray feedback mode mainly quenches residual star formation in massive galaxies. Our analysis reveals that this mode is also necessary to produce the first fully quenched galaxies before $z=2$, where the jets alone are inefficient. These initially highly star-forming galaxies contain relatively large black holes, likely strengthening the X-ray-powered heating and ejection of gas from the dense, central region of galaxies. Such extra heating source quenches the local star formation and produces a more variable accretion rate. More generally, this effect also causes the break down of correlations between the specific star formation rate, the accretion rate and the black hole mass. 
\end{abstract}

% Select between one and six entries from the list of approved keywords.
% Don't make up new ones.

%%%%%%%%%%%%%%%%%%%%%%%%%%%%%%%%%%%%%%%%%%%%%%%%%%

%%%%%%%%%%%%%%%%% BODY OF PAPER %%%%%%%%%%%%%%%%%%

\section{Introduction}
\label{sec:introduction}
Understanding the physical processes that shape the cosmic star formation history, and thus the stellar mass assembly in dark matter haloes, is essential to develop a comprehensive model of galaxy evolution.
While the hierarchical formation of dark matter structures is well established through the convergence of efficient gravity-only N-body simulations \citep{Frenk1988TheMatter, Springel2005TheGADGET-2,Primack2005PrecisionCosmology,Diemand2007FormationSubstructure,Klypin2011DarkSimulation,Angulo2012ScalingSimulation,Frenk2012DarkStructure}, the hydrodynamical interactions governing baryonic components are complex and costly to model computationally.

Early attempts to model the formation of stars from the cold halo gas either utilised fully analytic approaches \citep{White1978CoreClustering,Dekel1986TheFormation}, or semi-analytic models (earliest attempts by \citealt{White1991GalaxyClustering,Kauffmann1993,Cole_1994}), in which the evolution of baryons is modelled onto dark matter haloes extracted from N-body simulations. These works showed that assuming the cooling time as the bottleneck for star formation leads to an overproduction of stars with respect to observations. This is known as the `overcooling problem', which was subsequently confirmed by early hydrodynamical simulations \citep{Dave2001BaryonsMedium,Balogh2001RevisitingCrisis}. 

The solution to the overcooling problem lies in the inclusion of so-called `feedback processes', an umbrella term for self-regulating mechanisms arising from star formation and the supermassive black holes accreting in the centre of massive galaxies (active galactic nuclei; AGN). Pioneering works by \citet{Larson1974EffectsGalaxies} and \citet{Dekel1986TheFormation} showed that the momentum output from supernovae can cause significant gas-loss and thus deplete the reservoir for star formation. \citet{Silk1998LetterFormation} similarly argued that the release of gravitational potential energy from a fast-accreting AGN becomes comparable to the binding energy of the host galaxy, which may contribute to quenching. Modern analytic \citep{HS03, Rasera_2006, Dave_2012, Sharma2020TheFormation,Salcido_2020, Sorini2021ExtendedFormation, Fukugita_2022} and semi-analytic \citep[][]{Henriques_2015,Hirschmann2016GalaxyModel,Lacey_2016} models include at least some of these feedback processes and generally predict a more realistic star formation history. 

Nevertheless, the exact quenching mechanisms, their relative importance, and the coupling of the released energy to the interstellar gas, remain unclear (see reviews by \citealt{Somerville2015PhysicalFramework,Naab2016TheoreticalFormation}). A range of physical processes acting on different temporal and spatial scales may contribute to the overall quenching of galaxies. Full hydrodynamic cosmological simulations are thus a valuable testing ground for different theories, as increased computational power and refined hydrodynamics code solvers (\citealt{Hernquist1989TREESPH:Method,Teyssier2002CosmologicalRAMSES,Springel_2005, Gadget4, Schaller2023Swift:Applications}) have enabled the modelling of all galactic components across cosmic time. Despite this progress, numerical constraints often require a trade-off between resolution and simulated volume, hence making it unfeasible to describe sub-galactic feedback process within cosmological simulations from first principles. Instead, their presumed macroscopic effects are modelled via so-called sub-grid prescriptions \citep[see e.g.][for a recent review]{Crain_2023}. 

The general consensus is that stellar feedback is primarily efficient at quenching low-mass systems, due to their relatively low gravitational potential, and at early times, where AGN have not yet grown and matured enough to affect the galaxy. Associated processes such as supernovae and stellar winds are largely understood to cause outflows, in line with observations of high mass-loss rates in star-forming galaxies \citep{Heckman1990OnSuperwinds,Veilleux2005GalacticWinds,Rubin2014Evidence0.5}. High cooling rates in the dense environments at high redshift mean that most thermal energy input cools away too quickly to have a lasting effect on the ISM \citep{Somerville2015PhysicalFramework}. 

At late times, AGN are thought to deposit significant amounts of thermal and kinetic energy into the gas reservoirs of massive galaxies. Observational studies have broadly identified two main regimes. At accretion rates close to the Eddington limit, AGN feedback is radiatively efficient. Gravitational potential energy from the infalling matter is released as radiation, triggering winds capable of driving cool ISM gas out of the galaxy at non-relativistic speeds \citep[e.g.][]{Simionescu2008Metal-richXMM-Newton,Rupke2011Integral231,Fabian2012ObservationalFeedback,Cicone2014MassiveObservations, Harrison2014Kiloparsec-scalePopulation,King2015PowerfulNuclei,Kubo2022An3.09,Concas2022BeingHosts}. This is often called `quasar' or `wind' mode. The radiatively inefficient regime at low accretion rates is referred to as `jet' or `radio' mode, as these AGN can launch bipolar radio jets at relativistic speeds, carrying large amounts of kinetic and thermal energy \citep{Fabian2012ObservationalFeedback,Heckman2014TheUniverse,Blandford2019RelativisticNuclei}. Galaxies with observed AGN jets are characterised by low star formation rates, hinting at efficient quenching mechanisms. 

Some studies suggest that jets deposit their thermal energy into the circumgalactic medium (CGM) and intracluster medium (ICM). For example, jets often coincide with bubbles of hot gas visible in X-ray, which regularly penetrate as far as hundreds of kpc into the ICM and have likely been heated by shocks \citep{Suresh_2015,Turner_2017,Mukherjee2018RelativisticDiscs,Fielding_2020,Christiansen_2020}. This could be enough to offset cooling flows, i.e. the accretion of cooled CGM gas onto the galaxy, and thus starve active star formation \citep{Croton2006TheGalaxies,Gabor2015HotQuenching,Peng2015StrangulationGalaxies}. In some instances, jet-driven outflows have been found to be significantly mass-loaded to limit star formation via direct removal of gas from the central regions \citep{Morganti2013RadioAction,Jarvis2019PrevalenceQuasars}. Jets could also heat and accelerate existing wind-driven outflows \citep{Tadhunter2014JetIC5063,Dasyra2014Heating4C12.50}, potentially boosting their efficiency in quenching star formation. In addition to AGN-driven winds and jets, X-ray radiation from the accretion disk is thought to heat and exert pressure on the gas surrounding the black hole \citep{Sazonov2004Quasars:Heating,Hambrick2011TheEvolution,Choi2012RadiativeCode}. 

State-of-the-art cosmological simulations typically incorporate both stellar and AGN feedback mechanisms. Some simulations adopt a single AGN feedback prescription \citep[e.g.][]{Schaye2015TheEnvironments, Dave2016MUFASA:Hydrodynamics, McCarty_2017, Flamingo}, while others distinguish between a radiative wind mode at high accretion rates and a jet mode at low accretion rates \citep[e.g.][]{Vogelsberger2013APhysics, Dubois_2014, Vogelsberger2014IntroducingUniverse, Pillepich2018SimulatingModel, Dave2019Simba:Feedback,  MTNG}. Thus, the exact implementations of feedback processes differ substantially between simulations and rely on distinct physical assumptions. Nevertheless, they all broadly succeed at reproducing several observational benchmarks, such as the galaxy stellar mass function \citep[e.g.][]{Baldry_2008, Baldry_2012, Bernardi_2013, D_Souza_2015}, the star formation efficiency \citep[e.g.][]{Guo_2011, Behroozi_2013, Moster_2013}, the galaxy black-hole-stellar-mass relationship \citep[e.g.][]{Kormendy_2013, McConnell_2013}, or the gas fraction within haloes \citep[e.g.][]{Giodini_2009, Lovisari_2015}, although in some cases free parameters were tuned to obtain this agreement.

While encouraging, the broad agreement between observations and different simulations makes it harder to discern the validity and limitations of different models of galaxy formation. However, activating only specific feedback modes across different variants of otherwise identical simulations can reveal the exact effect of each prescription on the star formation history. Applying this strategy to the \textsc{Owls} project \citep{OWLS,van_de_Voort_2011} showed that AGN feedback is a major cause for the decline of the cosmic star formation rate density after $z=2$. Qualitatively similar conclusions were reached by \citet{Weinberger2017SimulatingFeedback} and \citet{Salcido2018TheHold} using different variants of the \textsc{IllustrisTNG} \citep{Pillepich2018SimulatingModel} and \textsc{Eagle} \citep{Schaye2015TheEnvironments} simulations, respectively. 

A rigorous study seeking to isolate the effect of every feedback prescription on star formation history has not yet been undertaken for the \textsc{Simba} simulation \citep{Dave2019Simba:Feedback}, which encompasses key novelties in the modelling of AGN feedback and black hole accretion with respect to previous state-of-the-art cosmological simulations. On top of implementing both jet and radiative wind feedback kinetically, \textsc{Simba} is the first cosmological-scale simulation to also explicitly model thermal energy input from X-ray heating processes, and to adopt a dual gas accretion mechanism onto black holes depending on the gas temperature. The simulation suite contains five variants of a $50 h^{-1} \mathrm{cMpc}$ cosmological box, which isolate \textsc{Simba}'s four different feedback prescriptions. These have already been used to constrain the effects of different feedback modes on the global statistics of galaxy properties \citep{Dave2019Simba:Feedback}, the distribution and thermal state of baryons within haloes and in the intergalactic medium \citep{Borrow2020CosmologicalSimulations,Sorini2022, Khrykin_2024}, the circumgalactic medium \citep{Appleby2019TheSimulation} and galaxy profiles \citep{Appleby2021TheSimba}.

In this work, we provide a comprehensive theoretical study on how \textsc{Simba}'s stellar and AGN feedback modes shape the cosmic star formation history. First, we investigate the signature of each feedback mode on global observables, such as the cosmic star formation rate density, on different halo populations, and even individual galaxies. We then connect our results with key properties of the simulated galaxies and their central black holes. 

The paper is organised as follows. In \S~ \ref{sec:simulations}, we describe the \textsc{Simba} runs considered. We then present our analysis, following a global-to-local approach. In \S~\ref{sec:global_evolution}, we assess the overall evolution of cosmic star formation by examining global star formation and stellar mass statistics in the five simulation runs. In \S~\ref{sec:mass dependence}, we examine the star formation history within halo populations of different mass. In \S~\ref{sec:AGN feedback z2}, we focus on the contribution to quenching from different AGN feedback modes on individual galaxies after $z \sim 2$, while \S~\ref{sec:xray} specifically focuses on effects caused by the X-ray heating mode on highly star-forming galaxies and black hole growth. In \S~\ref{sec:discussion}, we coalesce our results into a cohesive picture of how the different feedback prescriptions shape the cosmic star formation history and further compare our results with previous works on \textsc{Simba} as well as other simulations. Finally, \S~\ref{sec:conclusions} summarises our results. 

\section{Simulations}
\label{sec:simulations}
\subsection{\textsc{Simba}}
\textsc{Simba} \citep[described in detail in][]{Dave2019Simba:Feedback} is a suite of cosmological simulations built upon the \textsc{Gizmo} gravity plus hydrodynamics solver \citep{Hopkins2015AMethods}. \textsc{Gizmo} is composed of the \textsc{Gadget}-3 tree-particle-mesh gravity solver \citep{Springel2005TheGADGET-2} for dark matter particles and a meshless finite mass approach for the hydrodynamic evolution of gas particles. The simulations assume a cosmological model according to the \citet{Planck_2016} specifications, with the following parameters: $\Omega_M=0.3, \Omega_{\Lambda}=0.7, \Omega_b=0.048, H_0=68 \mathrm{~km} \mathrm{~s}^{-1} \mathrm{Mpc}\,{h}^{-1} \sigma_8=0.82, n_s=0.97$. Cosmological initial conditions are created using \textsc{Music} \citep{Hahn2011Multi-scaleSimulations}.

Star formation in \textsc{Simba} is modelled according to a sub-grid implementation by \citet{Krumholz2011AGalaxies} of an $H_{2}$-based \citet{Schmidt1959TheFormation.} law, where the $H_{2}$ fraction is estimated using the local metallicity and column density of the gas. The star formation rate (SFR) is then computed using the ratio of the molecular gas density and dynamical time, with a star forming efficiency of 0.02, inspired by \citet{Kennicutt1998TheGalaxies}. If the hydrogen number density $n_{\mathrm{H}}$ is above the threshold $n_{\mathrm{th}} = 0.13 \mathrm{~cm}^{-3}$, and the temperature is at most 0.5~dex above
\begin{equation}
\label{eq:ISM}
\log \left(\frac{\mathrm{T}}{\mathrm{K}}\right)=4+\frac{1}{3} \log \frac{n_{\mathrm{H}}}{n_{\mathrm{th}}},
\end{equation} 
a gas element is defined as interstellar medium (ISM) gas and can spontaneously turn into a stellar particle, conserving mass and momentum. 

Type II supernovae (SNe), Type Ia SNe, and asymptotic giant branch (AGB) stars will release metals into the surrounding gas, according to the chemical enrichment model by \citep{Oppenheimer_2006}, which tracks the evolution of eleven different elements (H, He, C, N, O, Ne, Mg, Si, S, Ca, Fe). This model also includes the sequestration of metals into dust particles. Radiative cooling and photoionisation heating processes are incorporated into the simulation via the \textsc{Grackle}-3.1 library \citep{Smith_2017}. \textsc{Simba} also employs a uniform ionizing background as detailed in \citet{HM12}, with modifications made by \citet{Rahmati_2013} to incorporate self-shielding on the fly.

Feedback from star formation is modelled as a kinetic ejection of metal-enriched wind particles, representing the aggregate effect of Type II supernovae, stellar winds and radiation pressure. 
Gas elements in proximity to star forming regions are decoupled from the bulk flow and expelled with a new velocity perpendicular to their prior velocity and acceleration vectors. Winds are two-phase, meaning 30$\%$ of the wind particles are ejected in a ``hot'' state, with a temperature based on the supernova energy and the kinetic wind energy, while the remaining 70$\%$ are ejected at $10^3\rm K$. The mass-loading factor of the wind is scaled with the galaxy's stellar mass according to a broken power law, a fit to mass outflow rates in the FIRE zoom-in simulations \citep{Angles-Alcazar2017TheSimulations}:
\begin{equation}
\eta\left(M_{\star}\right) \propto\left\{\begin{array}{ll}
9\left(\frac{M_{\star}}{M_{0}}\right)^{-0.317}, & \text { if } M_{\star}<M_{0} \\
9\left(\frac{M_{\star}}{M_{0}}\right)^{-0.761}, & \text { if } M_{\star}>M_{0}
\end{array}\right.,
\label{eq:massloading}
\end{equation}
where $\mathrm{M_{0}=5.2 \times 10^{9} \, M_{\odot}}$. \\
\textsc{Simba}'s wind speed scaling is based on the rates predicted by \citet{Muratov_2015}, also using the FIRE simulations. In order to represent enrichment due to Type II supernovae, wind metallicities are determined by the Type II supernovae yield and the mass-loading factor. Eventually, ejected particles will recouple, based on the surrounding gas conditions and the elapsed time since decoupling.

\textsc{Simba}'s black hole particles are seeded into galaxies with stellar mass $\mathrm{M_{\star}} \gtrsim 10^{9.5}\mathrm{\, M_{\odot}}$, with initial seed masses of $10^4 \, \mathrm{M}_{\odot}h^{-1}$. Growth occurs via two accretion modes, which are implemented as sub-grid models. They distinguish between a spherical Bondi accretion from hot non-ISM gas \citep{Bondi_1952} and a disk-like torque-limited accretion mode from cold ISM gas.
This mode is driven by gravitational instabilities and is able to recover the relation between black holes and their host galaxy masses without self-regulation
\citep{Hopkins_2011,Angles-Alcazar_2013,Angles-Alcazar_2015,Angles-Alcazar_2017a}.

\textsc{Simba} further employs three AGN feedback modes, which aim to model quenching mechanisms driven by the release of gravitational potential energy in the accretion process. These modes are active depending on the mass of the black hole, its accretion rate, and the gas fraction of the host galaxy. 
At high accretion rates with Eddington ratio $f_\mathrm{edd} > 0.2$, defined relative to the Eddington rate of the black hole particle, the radiative AGN wind mode produces bipolar outflows, which are modelled as a momentum injection in the direction parallel to the disk's angular momentum. Gas particles are randomly selected within the black hole kernel, decoupled and ejected at a new speed, given by:
\begin{equation}
\label{eq:winds}
v_{\mathrm{W}, \mathrm{EL}}=500+500\left(\log \mathrm{M_{BH}}-6\right) / 3 \mathrm{~km} \mathrm{~s}^{-1}, 
\end{equation}
where $\mathrm{M_{\mathrm{BH}}}$ is the black hole mass in solar masses $\mathrm{M_{\odot}}$. At top speed, the gas particles are ejected at 1200$\mathrm{~km} \mathrm{~s}^{-1}$. 

Below 0.2 $f_\mathrm{edd}$, AGN transition into jet mode, given their black hole mass $\mathrm{M_{\mathrm{BH}}}$ exceeds $10^{7.5}\mathrm{\, M_{\odot}}$. Similarly to the winds mode, this mechanism is modelled kinetically. Gas particles are ejected in the same fashion as before, but with a boosted velocity. This boost is inversely dependent on $f_\mathrm{edd}$:
\begin{equation}
\label{eq:jet}
v_{\mathrm{w}, \text { jet }}=v_{\mathrm{w}, \mathrm{EL}}+7000 \log \left(0.2 / f_{\mathrm{edd}}\right) \mathrm{~km} \mathrm{~s}^{-1}.
\end{equation}
The velocity increase due to jets is capped at 7000$\mathrm{~km} \mathrm{~s}^{-1}$. The fastest particles accelerated by AGN winds and jets can thus reach up to 8200$\mathrm{~kms}^{-1}$, an order of magnitude faster than in the pure radiative winds mode. While gas particles are ejected at the original temperature in the wind mode, jet mode outflows are heated to the virial temperature of the halo, $\mathrm{T_{vir}=10^{7}(M_{\text {H}} / 10^{15} \, M_{\odot})^{1 / 3} \mathrm{~K}}$ \citep{Voit2005ExpectationsRelations}, motivated by observations of hot gas carried in jets \citep{Fabian2012ObservationalFeedback}. The gas is recoupled after $10^{-4}$ of the Hubble time has passed.

Lastly, \textsc{Simba} accounts for heating from X-rays emitted by the accretion disk, which is quantified according to Equation 12 of \citet{Choi2012RadiativeCode}. \textsc{Simba}'s X-ray mode operates on galaxies with active jets, if the galaxy's gas fraction $f_\mathrm{gas} = \frac{M_\mathrm{gas}}{M_{\star}}$ is also below 0.2. In contrast to the wind and jet modes, energy input is spherical and impacts gas particles according to their proximity to the black hole. Non-ISM gas experiences an increase in temperature, while for ISM gas, half of the X-ray energy is converted into heat, and the remaining half is injected as radial outward momentum.

\subsection{Runs}
\label{sec:runs}
The \textsc{Simba} simulation suite comprises a range of simulation runs with different volumes and input physics. Unless otherwise specified, co-moving units are indicated with a `c' prefix. In this work, we make use of five simulation variants, which each span a volume of $50 h^{-1} \mathrm{cMpc}$ and are populated by $512^{3}$ dark matter and $512^{3}$ gas elements. They were performed from $z=249$ to $z=0$ with the same initial conditions, but sequentially deactivated feedback prescriptions, as outlined in  Table \ref{tab:runs}. Simba-50, the fiducial run, contains all four prescriptions, for a full physics simulation. No-X-ray excludes only the X-ray mode, whereas the No-jet run excludes both jets and X-ray heating. In the No-AGN run, the winds mode is deactivated as well, such that only the stellar feedback prescription remains. The No-feedback run straightforwardly contains no feedback processes. Contrasting results from these different runs will allow us to isolate the effects of each of the four different sub-grid prescriptions in the following analysis.

We further considered three additional runs of different volumes and resolution in order to test convergence. However, for these runs only the full fiducial setup is available and, consequently, we can only assess volume and resolution convergence for the Simba-50 run. Simba-25 and Simba-100 were performed at the same resolution as Simba-50, containing $2 \times 256^{3}$ and $2 \times 1024^{3}$ particles in volumes $25h^{-1}\mathrm{cMpc}$ and $100h^{-1}\mathrm{cMpc}$, respectively. Simba-25 High-res. contains $2 \times 512^{3}$ particles in its $25h^{-1}\mathrm{cMpc}$ volume. We note that in this run, the particle masses have been reduced by a factor of 8, to compensate for the increase in number of particles. A convergence test can be found in the appendix \ref{appendix:conv}.

Haloes were identified on the fly within \textsc{Simba}, using \textsc{Gizmo}'s 3D friends-of-friends (FOF) and a linking length of 0.2 times the mean inter-particle distance. FOF haloes were then cross-matched with haloes and galaxies found using the \textsc{Python} package \textsc{Caesar} \citep{Dave2019Simba:Feedback}. \textsc{Caesar} also generates catalogues containing all identified haloes and galaxies, along with key properties such as their masses and SFR's, which form the basis of the following analysis. The majority of these files are publicly available in the \textsc{Simba} Project Repository\footnote{http://simba.roe.ac.uk/}.

\begin{table*}\centering
\caption{\textsc{Simba} simulation runs used in this work.}
\begin{tabular}{@{}lcccccc@{}}

\toprule
\textbf{Simulation} & \textbf{Box size} ($h^{-1} \mathrm{cMpc}$) & \textbf{Nr. of particles} & \textbf{Stellar Feedback} & \textbf{Radiative winds}    & \textbf{Jets}         & \textbf{X-ray heating} \\ \midrule
Simba-100           & 100                             & $2 \times 1024^{3}$       & \cmark     & \cmark & \cmark & \cmark  \\
Simba-50            & 50                              & $2 \times 512^{3}$        & \cmark     & \cmark & \cmark & \cmark  \\
Simba-25           & 25                             & $2 \times 256^{3}$       & \cmark     & \cmark & \cmark & \cmark  \\
Simba-25 High-res.          & 25                             & $2 \times 512^{3}$       & \cmark     & \cmark & \cmark & \cmark  \\
No-X-Ray            & 50                              & $2 \times 512^{3}$        & \cmark     & \cmark & \cmark &                        \\
No-jet              & 50                              & $2 \times 512^{3}$        & \cmark     & \cmark &                       &                        \\
No-AGN              & 50                              & $2 \times 512^{3}$        & \cmark     &                       &                       &                        \\
No-Feedback         & 50                              & $2 \times 512^{3}$        &                           &                       &                       &                        \\ \bottomrule
\end{tabular}
\label{tab:runs}
\end{table*}

\section{Results}
\label{sec:results}
In the following analysis, we contrast key observables across the five $50 h^{-1} \mathrm{cMpc}$ simulation variants in order isolate the global effects of \textsc{Simba}'s stellar and AGN feedback modes on the cosmic star formation history. We subsequently connect them to the evolution of star formation and stellar mass in halo populations of different mass, as well as individual galaxies. 

\citet{Dave2019Simba:Feedback} used some of the observables presented in this paper to test \textsc{Simba}'s ability to reproduce key galaxy properties. However, thus far there has not been an analysis contrasting all five $50 h^{-1} \mathrm{cMpc}$ runs with different feedback prescriptions for a unified picture of how they affect \textsc{Simba}'s cosmic star formation history. Where appropriate, we provide a summary these prior assessments in the context of our results and refer the reader to the relevant publication for more detailed comparisons with observational results.

\subsection{The evolution of the global star formation history}
\label{sec:global_evolution}
We begin with an assessment of how \textsc{Simba}'s feedback modes affect the overall cosmic star formation and the resulting stellar mass growth. 

\subsubsection{Cosmic star formation rate density}
\label{sec:SFRD}
\begin{figure}
\centering
	% To include a figure from a file named example.*
	% Allowable file formats are eps or ps if compiling using latex
	% or pdf, png, jpg if compiling using pdflatex
	\includegraphics[width=\columnwidth]{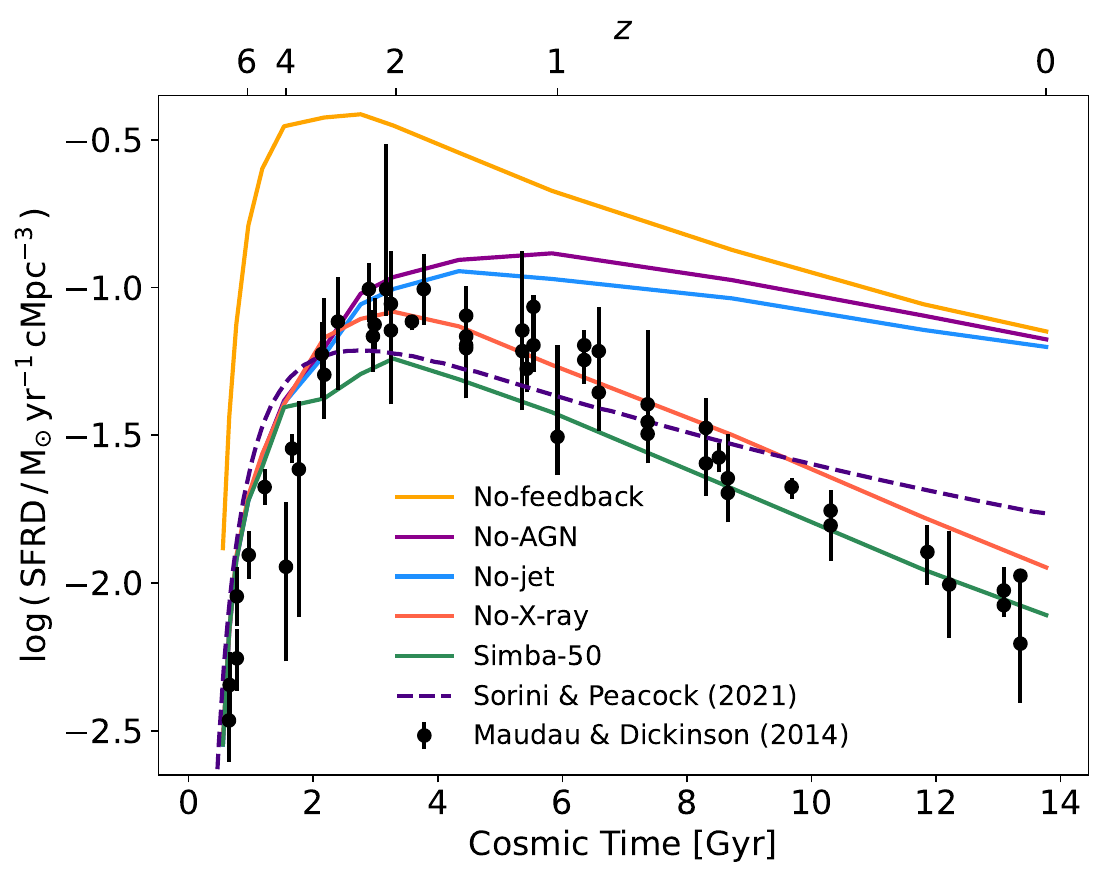}
    \caption{Cosmic star formation rate density for decreasing redshift, in $50 h^{-1} \mathrm{cMpc}$ \textsc{Simba} simulation variants with different feedback prescriptions (solid lines): No-feedback (yellow), No-AGN (purple), No-jet (red), No-X-ray (blue), and the full physics Simba-50 (green) run. Additionally shown are a collection of survey results \citep[data points]{Madau2014CosmicHistory} and an analytical model from \citet[indigo dashed line]{Sorini2021ExtendedFormation}. While stellar feedback is responsible for quenching the SFRD before $z \sim 2$, AGN feedback, specifically the jets, takes over as the dominant quenching mode thereafter. Including X-ray heating already produces a minor suppression from $z \sim 4$. }
    \label{fig:SFRD}
\end{figure}

The cosmic star formation rate density (SFRD) tracks the overall star formation rate of the simulation volume and is thus a good starting point for our analysis. In Fig. \ref{fig:SFRD}, we compare the temporal evolution of the SFRD across the five $50 h^{-1} \mathrm{cMpc}$ simulation variants. In order to compute the SFRD, we extract and sum up the SFRs for all FOF haloes at distinct snapshots between $z = 0-9$ and then divide each value by the comoving volume of the simulation box. We repeat this computation for the Simba-50 (green), No-X-ray (blue), No-jet (blue), No-AGN (purple), and No-Feedback (orange) runs. A collection of survey results by \citet[][black data points, hereafter MD14]{Madau2014CosmicHistory} is shown for comparison. 

The No-feedback run exhibits a sharp increase in SFRD before $z \sim 4$, culminating in a peak around $z \sim 3$ and a gradual decline at low redshifts. This conflicts with the observations from MD14, who find a later peak at $z \sim 2$, with its maximum value 0.7~dex below the No-feedback curve. The discrepancy between the predicted and observed SFRD grows up to $1\, \rm dex$ by $z=0$. The excess production of stars at all cosmic times in the No-feedback variant of \textsc{Simba} is an unsurprising manifestation of the aforementioned overcooling problem. However, within a hierarchical structure formation scenario, and in the absence of any feedback mechanism, the SFRD is expected to form more stars even earlier than in the No-feedback model considered here, due to its limited resolution, and therefore worsen the overcooling problem \citep{Dave2001BaryonsMedium}. The high-redshift growth of the SFRD in the No-feedback run is thus partly a numerical artefact. At higher redshift, massive haloes become increasingly rare due to the evolution of the halo mass function, hence a progressively larger fraction of haloes falls below the mass resolution limit of the simulation. Clearly, such haloes cannot contribute to the production of stars.

\textsc{Simba} runs with feedback modes are increasingly successful at reproducing the MD14 results. Including stellar feedback in the No-AGN run appears to efficiently quench the SFRD at high redshift. The slowed increase up to the peak at $z \sim 2$ is well represented, but it fails to reproduce the declining SFRD of the MD14 results thereafter. Including AGN radiative winds in the No-jet run, in addition to stellar feedback, suppresses the SFRD by less than 0.1 dex at $z < 2$ and has no discernible effect before. Only when including AGN jets, in the No-X-ray run, the SFRD starts showing a slight downturn just before $z=2$ and is strongly suppressed thereafter. The slope of the SFRD after cosmic noon becomes steeper, closely resembling the shape traced by the data. By redshift $z \sim 0$, the jets have produced a cumulative suppression of almost 1~dex. 

Interestingly, the impact of different feedback prescriptions on the SFRD that we just described is in broad agreement with the predictions of recent analytical models of cosmic star formation. 
We report in Fig.~\ref{fig:SFRD} the SFRD given by the \cite{Sorini2021ExtendedFormation} model (purple dashed line) as an example. Such model constitutes a generalisation of the formalism introduced by \cite{HS03} and calculates the SFRD by integrating the SFR within haloes of a given total mass, weighted by the halo mass function. For a given population of haloes of fixed mass, the SFR is determined by whichever time scale is the shortest: the cooling time at low redshift, and an average gas consumption time scale \cite[based on the reasoning in][]{SH03_SFR} at high redshift. Additionally, \cite{Sorini2021ExtendedFormation} introduced a simple constant stellar wind model to mimic the effect of stellar feedback in reducing the gas reservoir available for the production of stars at any given time. 

It is noteworthy that the \cite{Sorini2021ExtendedFormation} model agrees with the fiducial \textsc{Simba} run not only $z\gtrsim 3$, where stellar feedback appears to be the dominant driver of quenching, but also around cosmic noon. The increasing discrepancy in the slope of the SFRD at later times is due to the absence of an explicit AGN feedback mechanism in the \cite{Sorini2021ExtendedFormation} model. This also explains why the slope given by the analytical model matches the one found in the No-AGN \textsc{Simba} run. However, analytical SFRD at $z<2$ exhibits a lower normalisation, yielding a better match with data. This happens because, although \cite{Sorini2021ExtendedFormation} do not model AGN feedback, the underlying parameters of their formalism are still calibrated to reproduce observations of the Kennicutt-Schmidt \citep{Kennicutt1998TheGalaxies} and baryonic Tully-Fisher \citep{bTFR} relationships (\citealt{Genzel_2010} and \citealt{Lelli_2016}, respectively). On the contrary, no parameter is recalibrated within the \textsc{Simba} suite of simulations once a specific feedback module is deactivated. It is therefore expected that, for the same set of feedback mechanisms, the \textsc{Simba} results lie farther away from the data. Similar considerations can be made when comparing the analytical SFRD at low redshift with the results of the No-jet run, which do not significantly deviate from those of the No-AGN run. 

We therefore suggest that the inclusion of some form of AGN-jet mechanical feedback in analytical models along the lines of \cite{Sorini2021ExtendedFormation} can improve the match with observations. However, this might not be the case for other analytical models that rely on a different formalism. For instance, a smaller impact of AGN versus stellar feedback on the late-time SFRD has been highlighted also by the \cite{Salcido_2020} analytical model, which, unlike \cite{Sorini2021ExtendedFormation}, includes a phenomenological implementation of feedback from radiative AGN winds \citep[following][]{Benson_2010, Benson_2012}. If the \cite{Salcido_2020} AGN prescriptions is turned off, the slope of the late-time SFRD remains essentially unchanged, while is still in agreement with observations. However, a comprehensive comparison of the numerical predictions of the different \textsc{Simba} variants against analytical models of star formation goes beyond the scope of this manuscript, and we leave it for future work.

Another result emerging from Fig.~\ref{fig:SFRD} is that X-ray heating may play a relatively important role in shaping the SFRD both at high and low redshift. Such feedback mechanism is included in the fiducial Simba-50 run and the resulting SFRD displays a further suppression of around 0.2~dex, in excellent agreement with MD14 at late times. However, it appears that before $z \sim 1$, the data are better reproduced by the No-X-ray run than the full-physics Simba-50 run. Nevertheless, the predictions of both the No-X-ray and Simba-50 runs agree with most data points within the error bars, hence current SFRD observations cannot definitively discriminate between these models. Furthermore, some data may be affected by systematic errors, since parametric models estimating the SFRD from the spectral energy distributions of observed galaxies may impose strong priors on the underlying physical parameters, hence biasing the results \citep{Carnall_2019}. Thus, the imperfect match between the No-X-ray or Simba-50 runs and the \cite{Madau_2014} compilation in certain redshift ranges should not be concerning. 

We further note that while quenching of the SFRD in the No-Xray and No-jet runs set in just before $z \sim 2$, the suppression in the Simba-50 run begins already at $z \sim 4$. We discuss the origin of this feature further in \S~\ref{sec:xray_SFR}. Lastly, we note that our curve for the Simba-50 run is consistent with \citet{Dave2019Simba:Feedback}, who in their Fig. 7 compare the SFRD in the full physics Simba-100 run to the MD14 results. This means that the predictions of the \textsc{Simba} variants displayed in Fig.~\ref{fig:SFRD} are well converged with respect to the box size (see also Fig. \ref{appendix:conv} and the discussion in appendix \S~\ref{app:A}). 

In summary, we conclude that \textsc{Simba}'s stellar feedback mode is responsible for quenching the SFRD before $z \sim 2$. Simulation variants including AGN feedback prescriptions do not exhibit a significant difference to the No-AGN run before $z \sim 2$, with the exception of a drop in SFRD of around 0.2~dex after $z \sim 4$ in the fiducial Simba-50 run. This indicates that, overall, AGN feedback is only able to efficiently quench cosmic star formation activity after $z \sim 2$, while the inclusion of X-ray heating creates a minor suppression already at $z > 2$. After $z \sim 2$, the AGN jet mode appears to be by far the most efficient at suppressing the SFRD, while the inclusion of X-ray heating only provides a minor contribution to quenching. Radiative AGN winds have almost no effect on the overall evolution of the SFRD.

\subsubsection{Galaxy stellar mass function}
\label{sec:GSMF}
\begin{figure*}
	% To include a figure from a file named example.*
	% Allowable file formats are eps or ps if compiling using latex
	% or pdf, png, jpg if compiling using pdflatex
	\includegraphics[width=\textwidth]{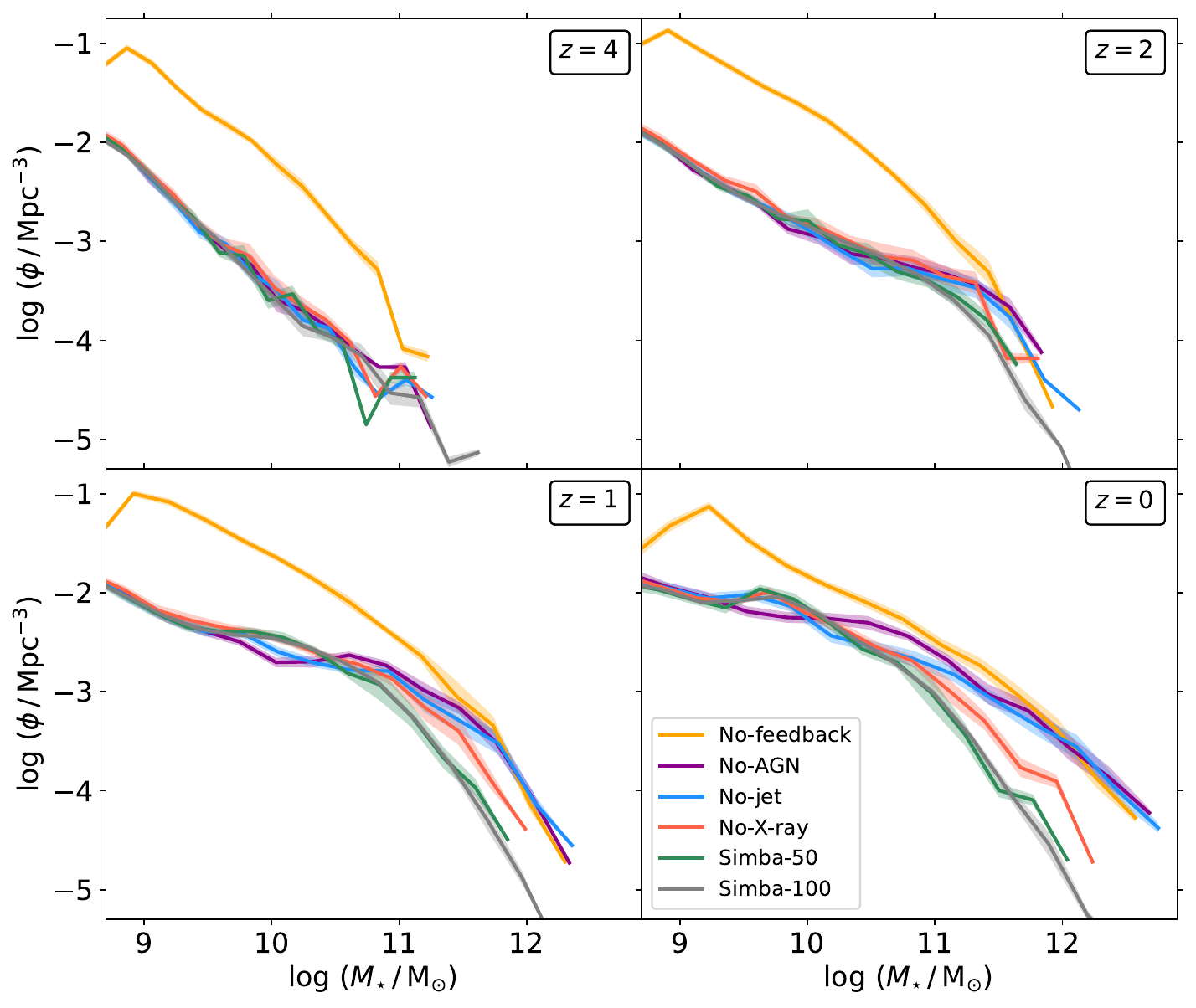}
    \caption{Galaxy stellar mass function (GSMF, here $\Phi$) in five simulation variants, shown in the same colours as in Fig. \ref{fig:SFRD}. Additionally shown is the Simba-100 run (grey). Errors (shaded areas) are computed by jackknife re-sampling. Overall, the GSMF is lowered by stellar feedback up to $z = 2$. Thereafter, AGN feedback suppresses the GSMF at stellar masses above $10^{10.5}\mathrm{\, M_{\odot}}$. Jets and X-ray heating also truncate the GSMF at the high-mass end at $z = 1$ and 0.}
    \label{fig:GSMF}
\end{figure*}

The effects described in \S~\ref{sec:SFRD} represent the `macroscopic' impact of different feedback mechanisms on the star formation history, as the cosmic SFRD is a global quantity that incorporates the contributions from all galaxies. In this section, we will investigate how the different feedback prescriptions affect the growth of galaxies of different stellar masses. This can be done by studying the galaxy stellar mass function (GSMF), which characterises the number density of galaxies per co-moving volume and logarithmic stellar mass bin. 
In Fig. \ref{fig:GSMF}, we present the GSMF (here denoted as $\Phi$) at $z = 4, 2, 1$, and $0$. In the same colour-code as in Fig. \ref{fig:SFRD}, we show the $50 h^{-1} \mathrm{cMpc}$ simulation variants, as well as the full-physics Simba-100 run (grey). Errors due to cosmic variance (shaded areas) have been estimated by jackknife re-sampling over the eight sub-octants of the simulation volume.

At $z=4$, all $50 h^{-1} \mathrm{cMpc}$ simulation runs are populated with galaxies containing stellar masses lower than $10^{11.25}\mathrm{\, M_{\odot}}$, and considerably more low-mass galaxies of around $10^{9}\mathrm{\, M_{\odot}}$ than higher-mass galaxies. Across all mass bins, the No-Feedback GSMF is around 1~dex higher than in all other runs. The runs with at least stellar feedback
appear almost identical, consistent with the result from \S~\ref{sec:SFRD} that stellar feedback is primarily responsible for quenching galaxies at early times, where AGN feedback effects are not yet significant. 

At $z = 2$, the GSMF flattens above $10^{11}\mathrm{\, M_{\odot}}$ for all runs containing feedback. They only diverge around  $10^{11.5-11.7} \mathrm{\, M_{\odot}}$, where the the No-X-ray and Simba-50 runs show a slight trend toward fewer high mass galaxies. However, at this epoch, all runs contain few massive galaxies, meaning the high-mass end of the GSMF is affected by small number statistics. There appears to be no significant difference between the No-X-ray and Simba-50 runs, indicating that the gap between the SFRD given by these runs at $z \sim 2-4$ (see Fig. \ref{fig:SFRD}) does not seem to translate into significant differences in the GSMF. This suggests that the early quenching due to the activation of the X-ray mode is not enough to slow the overall stellar mass growth.

If the stellar feedback mode is active, the low-mass end of the GSMF flattens out even more at $z = 1$ and 0, with a drop-off of varying steepness above $10^{10.5}\mathrm{\, M_{\odot}}$. We note that above this threshold the GSMF seems to converge across the No-feedback, No-AGN and No-jet runs. Both stellar feedback and AGN winds do not appear to be effective at suppressing the formation of galaxies with high stellar mass content. In contrast, the No-X-ray run and Simba-50 runs, show a decreased GSMF at these masses, indicating a suppression due to the AGN jets and the X-ray mode. Galaxies with stellar mass above $10^{12.25}\mathrm{\, M_{\odot}}$ and $10^{12}\mathrm{\, M_{\odot}}$ do not form in the No-X-ray and Simba-50 run, respectively. 

A minor peak emerges around $10^{9.5}\mathrm{\, M_{\odot}}$ for all runs containing AGN feedback, which coincides with the stellar mass necessary to seed a black hole particle into the galaxy (see \S~\ref{sec:simulations}). This peak reflects an accumulation of galaxies reaching this stellar mass and immediately being affected by their new black hole, which slows the growth of additional stellar mass. Except for this peak, the AGN feedback runs show no significant difference to the No-AGN run below $10^{10.5}\mathrm{\, M_{\odot}}$, indicating that the upper threshold of the GSMF is set by the stellar feedback mode in this mass regime. Lastly, we note that with decreasing redshift the low mass end of the No-feedback GSMF shrinks at the expense of a growing high mass tail. Thus, this run assembles most of its mass earlier than the other runs and primarily grows low mass galaxies to higher masses at late times. 

In conclusion, we find that before $z\sim 2$, the overall stellar mass growth in \textsc{Simba} is regulated by stellar feedback. After $z\sim 2$ the AGN jet mode, and to a lesser extent the X-ray mode, suppress the formation of galaxies with high stellar masses. Below $10^{10.5}\,\mathrm{M_{\odot}}$, the shape of the GSMF continues to be largely set by stellar feedback. 

Across the redshift range considered in Fig.~\ref{fig:GSMF}, the GSMF from the fiducial Simba-50 and Simba-100 runs are consistent with each other, indicating that our results are converged volume wise. A comparison of the GSMF against observations was already undertaken for the Simba-100 simulation across the redshift range $0<z<6$ by \cite{Dave2019Simba:Feedback}. The Simba-100 simulation was found to yield good agreement with data below $10^{11}\mathrm{\, M_{\odot}}$. However, similarly to other cosmological simulations \citep[e.g. \textsc{Eagle},][]{Schaye2015TheEnvironments}, \cite{Dave2019Simba:Feedback} identified an excess of galaxies at $10^{11}\mathrm{\, M_{\odot}}$. They attributed this to both numerical uncertainties and observational difficulties in quantifying massive systems, as well as dust-obscuration potentially causing restframe-optical surveys to miss massive galaxies. We refer the interested reader to \cite{Dave2019Simba:Feedback} for further details.

\subsection{Stellar- and halo-mass dependence of stellar and AGN feedback effects}
\label{sec:mass dependence}
In \S~ \ref{sec:global_evolution}, a consistent picture emerged for the global evolution of the cosmic star formation activity and the resulting stellar mass-buildup in \textsc{Simba}: stellar feedback is responsible for quenching the overall star formation before $z \sim 2$, whereas AGN feedback is the dominant mechanism at lower redshift. The evolution of the GSMF indicates that this transition is mass-dependent, where stellar feedback slows the stellar mass assembly in low-mass systems, but does not affect the formation of high-mass galaxies significantly. The number density of these galaxies is instead set by AGN feedback, in particular the jet feedback mode. In this section, we split \textsc{Simba}'s halo population according to mass to further examine how their star formation rates and relative stellar content are affected by each of the implemented feedback models. 

\subsubsection{Evolution of the star formation rate in low- and high-mass haloes}
\label{sec:halobins}
\begin{figure*}
	% To include a figure from a file named example.*
	% Allowable file formats are eps or ps if compiling using latex
	% or pdf, png, jpg if compiling using pdflatex
	\includegraphics[width=0.9\textwidth]{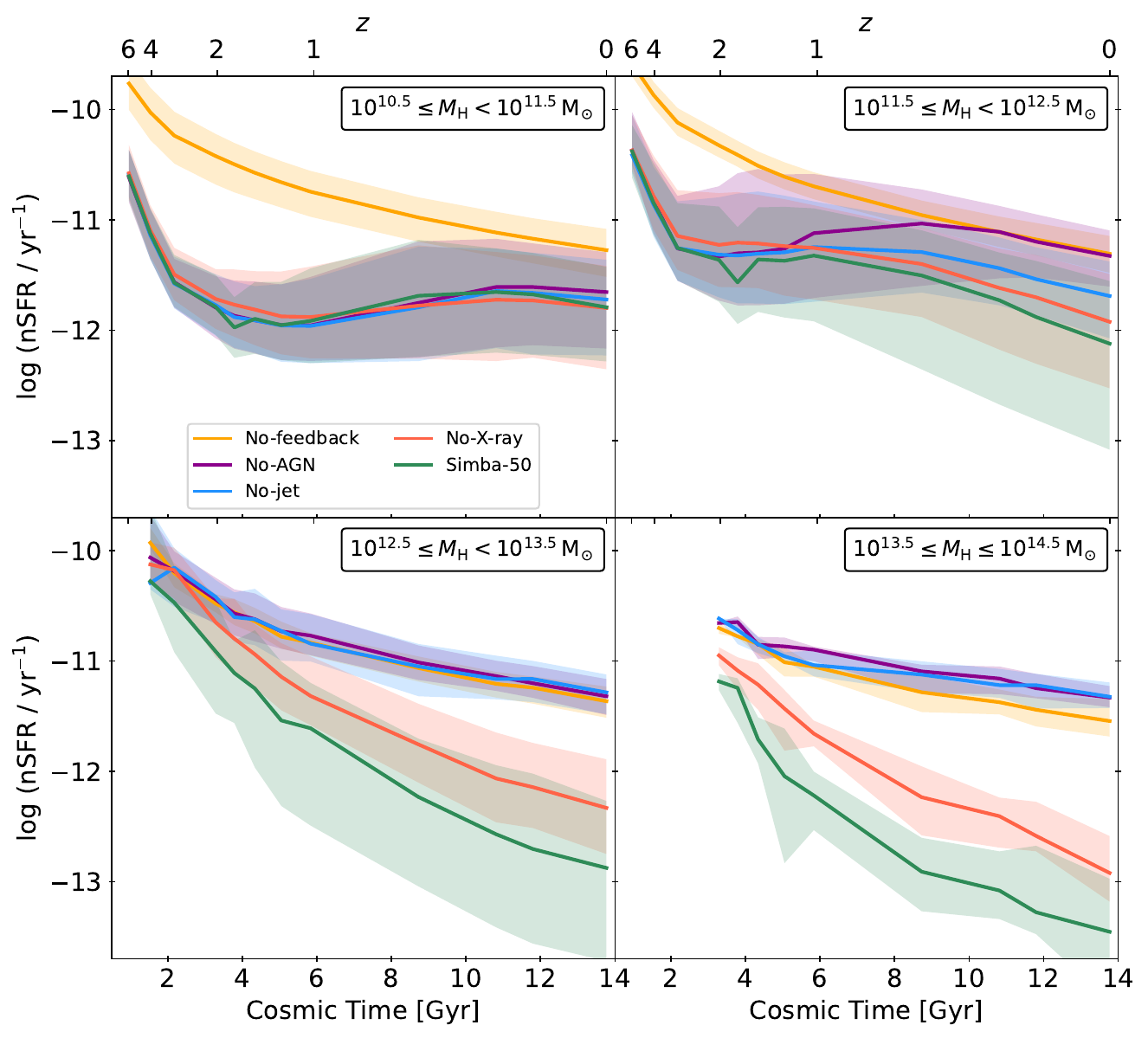}
    \caption{Median of normalised star formation rate (nSFR) with $\mathrm{16^{th}-84^{th}}$ percentile spread in four halo mass bins centred on $M_\mathrm{H} = 10^{11}\mathrm{\, M_{\odot}}$, $10^{12}\mathrm{\, M_{\odot}}$, $10^{13}\mathrm{\, M_{\odot}}$, and $10^{14}\mathrm{\, M_{\odot}}$, shown for simulation runs with different feedback implementations. Lower halo masses $\lesssim 10^{12}\mathrm{\, M_{\odot}}$ are mainly affected by stellar feedback, while AGN feedback quenches the nSFR at higher masses $\gtrsim 10^{13}\mathrm{\, M_{\odot}}$. The AGN jets emerge again as the most powerful mode.}
    \label{fig:halobins}
\end{figure*}

We first consider how the instantaneous star formation rate evolves in haloes of different total mass. At snapshots between $z=6$ and 0 in each of the $50 h^{-1} \mathrm{cMpc}$ simulations, we categorise all FOF haloes with at least one galaxy into four logarithmic mass bins. These bins are symmetric and centered on $M_\mathrm{H} = 10^{11}, 10^{12}, 10^{13}$, and $10^{14}\mathrm{\, M_{\odot}}$. For all haloes in the simulation variants, we find the normalised SFR ($\rm nSFR \equiv SFR/\mathit{M}_\mathrm{H}$), which provides a relative measure of the star forming activity for each halo. In Fig. \ref{fig:halobins}, we then compare the temporal evolution of the median nSFR in each of the mass bins across the simulation variants, colour-coded as in Fig. \ref{fig:SFRD} and \ref{fig:GSMF}. The shaded area represents the $\rm 16^{th}$ to $\rm 84^{th}$ percentile spread. 

For haloes in the $10^{11}\mathrm{\, M_{\odot}}$ bin, only the No-Feedback run differs from the other four simulation variants. The nSFR is up to 1.5~dex greater and exhibits a gradual decrease from $10^{-10} \, \mathrm{yr}^{-1}$ to just below $10^{-11} \, \mathrm{yr}^{-1}$. This evolution is similar at all halo masses, as the star formation limit is only set by the cooling time and the availability of cold and dense ISM gas.
Except for a very slight divergence close to $z\sim 0$, the nSFRs in the other simulation variants appear almost identical to the No-AGN feedback run, regardless of the modes activated. The median nSFR exhibits an initial quenching between redshifts 6 and 2 and subsequently continues at a roughly constant rate between $10^{-11.5} \, \mathrm{yr}^{-1}$ and $10^{-12} \, \mathrm{yr}^{-1}$. At these low masses, either black holes have not been seeded due to similarly low stellar masses, or if they have, they are in early evolutionary stages. This would mean that either the more powerful AGN jets and X-ray heating modes have not been active or their effect on the surrounding gas has not accumulated enough to exhibit a significant impact on the nSFR evolution. 

At $M_{\rm H} \sim 10^{12}\mathrm{\, M_{\odot}}$, the four simulations with feedback show differences after $z\sim 1$, where potential black hole particles in the more massive haloes within this bin have had time to be seeded and evolve. At this epoch, as well as across the entire evolution in all higher mass bins, the No-AGN run converges with the No-feedback run, while the runs with AGN feedback show a continued suppression.

\textsc{Simba}'s stellar feedback thus shapes the nSFR evolution in lower-mass haloes, but is not effective at higher masses. Above $M_{\rm H} \sim 10^{12.5}\mathrm{\, M_{\odot}}$, the stronger gravitational potential likely keeps accelerated gas elements from escaping the galaxy through outflows. Additionally, the mass loading of stellar winds follows a broken power law (see Eq. \ref{eq:massloading}). Haloes containing more stellar mass than the threshold value of $\mathrm{M_{0}=5.2 \times 10^{9} \, M_{\odot}}$, are assigned a mass loading factor that declines much more steeply with stellar mass. This reduced mass-loading factor results in fewer ejected gas particles. As larger haloes typically contain more stellar mass, this further diminishes the impact of stellar feedback in these larger systems. 

As for the effects of AGN feedback after $z\sim 1$ in the $M_{\rm H} = 10^{12}\mathrm{\, M_{\odot}}$ bin, we note that each feedback mode contributes to the overall quenching of the median nSFR. This is the only mass range, where the No-jet run shows a visible difference from the No-AGN run, meaning that radiative AGN winds have quenched the nSFR by around $0.5 \,\rm dex$ at $z = 0$. Similarly as for the stellar feedback, the gravitational potential of more massive haloes is likely too strong for wind-ejected particles to escape. Even for the largest black holes, the maximum speed a gas particle can reach in wind mode is around 1200 $\rm km\,{s}^{-1}$, while the escape velocity for massive haloes is approximately an order of magnitude greater. Therefore, \textsc{Simba}'s AGN-driven winds are only able to quench the star formation in haloes of around $10^{11.5}$ to $10^{12.5}\mathrm{\, M_{\odot}}$.

As a result, the influence from stellar and radiative wind feedback modes is negligible in the $10^{13}\mathrm{\, M_{\odot}}$ and $10^{14}\mathrm{\, M_{\odot}}$ mass bins. Instead, the AGN jets and X-ray heating modes are the dominant quenching mechanisms. At $z \sim 0$, the No-X-ray run shows a significant suppression of the median nSFR by approximately 1 dex in the $10^{13}\mathrm{\, M_{\odot}}$ bin and 1.5 dex in the $10^{14}\mathrm{\, M_{\odot}}$ bin. This confirms that \textsc{Simba}'s AGN jets are in large part responsible for the overall quenching of star formation in high mass haloes. In the Simba-50 run, the additional X-ray heating causes a further reduction of the median nSFR by $\sim 0.5 \, \rm dex$ in both high-mass bins. In the $10^{13}\mathrm{\, M_{\odot}}$ bin, the divergence from the No-X-ray run, appears as early as $z \sim 4$, consistent with the early quenching of the star formation rate density in the Simba-50 run seen in Fig. \ref{fig:SFRD}. In the $10^{14}\mathrm{\, M_{\odot}}$ bin, all nSFR curves begin at $z \sim 2$, meaning that such massive haloes have not formed before this epoch. The particular influence of the jet and X-ray feedback modes at these high halo masses is likely a consequence of larger-mass haloes housing black holes that are not only more massive and have a larger sphere of influence, but are also further along in their evolutionary trajectory. Older black holes tend to reach lower accretion rates, which will allow the activation the jet mode and, if the gas fraction is low, the X-ray mode. The jet mode, in particular, will significantly increase the ejection speeds and thermal energy of the wind particles. 

In summary, we find a clear mass dependence in how \textsc{Simba}'s feedback modes affect the evolution of the normalised SFR in different halo populations. Stellar feedback sets the star formation activity for haloes smaller than $10^{12.5}\mathrm{\, M_{\odot}}$. After $z \sim 1$, haloes between $10^{11.5}$ and $10^{12.5}\mathrm{\, M_{\odot}}$ are also affected by radiative winds, jets and X-ray heating. Above this threshold, stellar and AGN winds cease to be effective and instead the jets and X-ray heating modes become the significant quenching mechanisms in haloes above $10^{12.5} \mathrm{\, M_{\odot}}$. This mass dependence arises due to the competing factors from each system's gravitational potential, the evolutionary state of the AGN, as well as the mass-dependent parameter scaling of the stellar mode efficiency.  

\subsubsection{The stellar mass-halo mass ratio at $z \leq$ 2}
\label{sec:SMHM}
The previous section has shown that the AGN-jet and X-ray feedback modes in \textsc{Simba} are efficient at suppressing the median nSFR of haloes with $M>10^{12.5}\mathrm{\, M_{\odot}}$ and contribute to quenching haloes between $10^{11.5}$ and $10^{12.5}\mathrm{\, M_{\odot}}$ after $z \sim 1$. However, this broad statistic does not demonstrate how each mode shapes the stellar content of haloes as a result. To disentangle the impact of each of the three AGN feedback mechanisms, we thus examine the stellar mass-halo mass (SMHM) ratio, revealing the efficiency with which haloes of different mass turn their gas into stellar mass.

\begin{figure*}
	% To include a figure from a file named example.*
	% Allowable file formats are eps or ps if compiling using latex
	% or pdf, png, jpg if compiling using pdflatex
	\includegraphics[width=\textwidth]{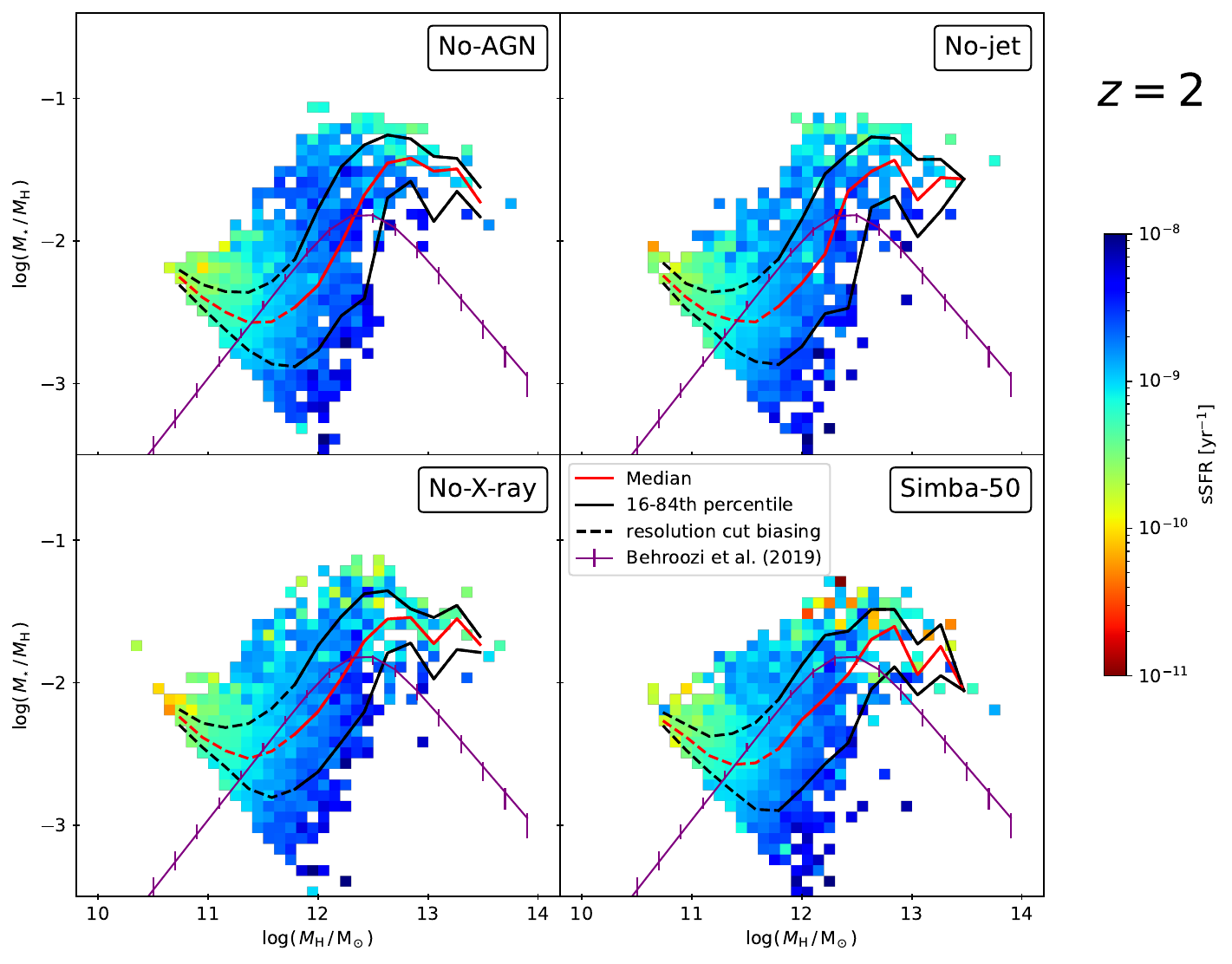}
    \caption{2-dimensional histograms of the SMHM ratio for $z = 2$ in the four simulation variants containing feedback. Bins are coloured by the sSFR. Additionally shown are the median relation (red lines) with the $\rm 16^{th}$ to $\rm 84^{th}$ percentile (black lines), as well as the \textsc{UniverseMachine} median \citep[purple]{Behroozi_2019}. Below halo masses of $\sim 10^{11.7}\mathrm{\, M_{\odot}}$ we expect the resolution cut to affect the median relation significantly (change from solid to dashed lines). At this epoch, star formation rates are generally high and all feedback runs exhibit remarkably similar features, except for a few quenched bins in the Simba-50 run due to X-ray feedback. This means, stellar feedback is the dominant mechanism shaping stellar mass assembly and AGN feedback is not yet efficient.}
    \label{fig:heatmapz2}
\end{figure*}

\begin{figure*}
	% To include a figure from a file named example.*
	% Allowable file formats are eps or ps if compiling using latex
	% or pdf, png, jpg if compiling using pdflatex
	\includegraphics[width=\textwidth]{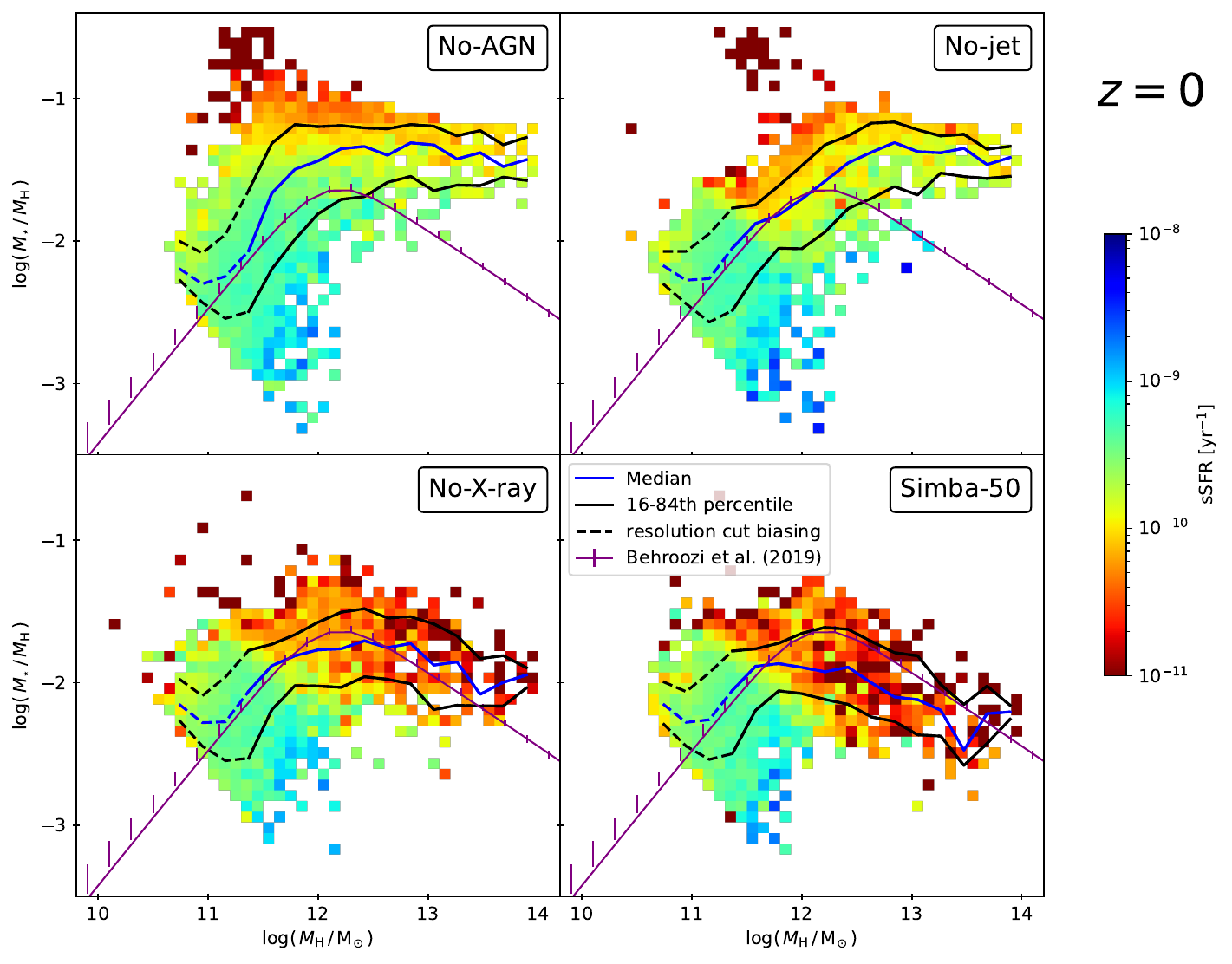}
    \caption{2-dimensional histograms of the SMHM ratio for $z = 0$, in a configuration otherwise identical to \ref{fig:heatmapz2}. The resolution cut biasing affects halo masses below $\sim 10^{11.2}\mathrm{\, M_{\odot}}$. For readability purposes, the median relation is shown in blue instead of red here. Jets introduce the sharp turnover in the No-X-ray run at $10^{12}\mathrm{\, M_{\odot}}$, showing a continuous suppression of stellar mass assembly in large haloes. AGN winds from newly seeded black holes suppress the sSFR at intermediate masses, while X-ray feedback contributes to the quenching of galaxies in massive haloes.}
    \label{fig:heatmapz0}
\end{figure*}

We calculate the SMHM ratios of all haloes at $z = 2$ and $z = 0$ by dividing the stellar mass within the central galaxy of each halo by the total halo mass. This calculation eliminates any contribution from small satellite galaxies, which might otherwise obscure the trends observed in the massive central galaxies that are most impacted by the growth and feedback of their central black holes. Fig. \ref{fig:heatmapz2} and Fig. \ref{fig:heatmapz0} then show two-dimensional histograms of the SMHM ratio against halo mass for the runs containing at least stellar feedback, respectively at $z = 2$ and $z=0$. In order to directly connect the relative stellar content to the star forming activity, the bins are coloured by the specific star formation rate $(\rm sSFR \equiv SFR/\mathit{M}_{\star}$) of the central galaxy. We additionally show the median SMHM ratio (red line in Fig. \ref{fig:heatmapz2}, blue line in Fig. \ref{fig:heatmapz0}) and $16^{\rm th}-84^{\rm th}$ percentile spread (black lines). At the low halo mass ends, the median relations are biased by the diagonal resolution cut-off, which enforces a minimum number of stellar particles for a halo to be considered well-resolved, corresponding to a stellar mass of $\sim 10^{8.5} \mathrm{\, M_{\odot}}$. This starts affecting the median relations below halo masses of approximately $10^{11.2} \mathrm{\, M_{\odot}}$ at $z=2$ and $\sim 10^{11.7} \mathrm{\, M_{\odot}}$ at $z=0$ (relations shown in dashed instead of solid lines). Above the resolution cut, we compare to the median from the \textsc{UniverseMachine} abundance matching \citep[][purple]{Behroozi_2019}, which assumes a one-to-one correspondence between galaxies and dark matter haloes by comparing observed galaxy properties to the properties of simulated haloes. 

% Example figure

At $z = 2$, the histograms exhibit a largely similar pattern across the shown runs, in agreement with the findings that \textsc{Simba}'s AGN feedback is largely active after $z=2$. Most bins show a high sSFR ranging from approximately $10^{-9}$ to $10^{-8} \, \mathrm{yr}^{-1}$. The lower end of the halo mass spectrum, below $10^{11}\mathrm{\, M_{\odot}}$, displays a moderate reduction in sSFR, likely due to the highly effective stellar feedback in small systems. In the full Simba-50 run, a few high-mass bins appear quenched, with specific star formation rates as low as $10^{-10}$ to $10^{-11} \, \mathrm{yr}^{-1}$. Moreover, the histogram in the Simba-50 run is slightly truncated, with the highest ratios reaching only up to $\sim 10^{-1.25}$ rather than extending to $\sim 10^{-1.10}$ as seen in the other runs. These haloes thus contain a relatively smaller fraction of stellar mass compared to haloes of the same mass in the other runs. This indicates that a minor suppression of stellar mass growth has taken place before $z = 2$, and is likely correlated with the early quenching of the star formation rate density in the Simba-50 run (see \S~ \ref{sec:SFRD}). We further find excellent agreement of the median relation with \citet{Behroozi_2019} in all feedback simulation runs, until the turnover at around $M_\mathrm{H}=10^{12.5}\mathrm{\, M_{\odot}}$. As shown in Fig. \ref{fig:halobins}, within this mass range the nSFR is governed by the stellar feedback mode at $z>1$, explaining the lack of differences between them. 
At the high mass end, the \textsc{Simba} simulations generally predict SMHM ratios that are larger than the median relation reported in \citet{Behroozi_2019}. None of the runs fully captures the turnover point in the relation, with the Simba-50 run showing the closest approximation. This means that even with all AGN feedback modes, \textsc{Simba} could overestimate the amount of stellar mass contained in high mass haloes at $z=2$. As referenced in \S~\ref{sec:GSMF}, this excess of high-mass galaxies is discussed in \citet{Dave2019Simba:Feedback}. For Fig. \ref{fig:heatmapz2} in particular, we also note that the \textsc{UniverseMachine} median is itself derived from significantly processed data and thus subject to additional assumptions. As a result, the mismatch with the Simba-50 run is not overly concerning.

At $z = 0$ the histograms of the SMHM ratio exhibit stark differences. sSFR's in the No-AGN run are most quenched at high stellar mass fractions, likely because large amounts of the gas have already been turned into stars. AGN winds have additionally quenched the sSFR in the No-jet run for halo masses between $10^{11}\mathrm{\, M_{\odot}}$ and $10^{12}\mathrm{\, M_{\odot}}$, which has largely prevented stellar mass fractions of $10^{-1.5}$ and above. These represent haloes with stellar masses larger than the seeding threshold $\mathrm{M_{\star,BH}} = 10^{9.5}\mathrm{\, M_{\odot}}$, which are impacted by their new black hole. While radiative winds have shown only a minimal ability to quench star formation in prior figures, we show here that this is sufficient to affect stellar mass growth in intermediate mass haloes. We further note that the No-AGN and No-jet runs include a population of highly quenched haloes with stellar mass fractions exceeding $10^{-1}$. This is an artefact of the Friends-Of-Friends halo finder, where small galaxies in their own subhalo, but in close proximity to a large halo, are interpreted as central galaxies instead of satellite galaxies. These populations only occur in the No-AGN and No-jet run, as the inclusion of jets prevents these galaxies from forming in the first place. We do not consider these populations true quenched central galaxies and thus exclude them from our further discussion. 

In the No-X-ray run, the jets introduce the expected turnover at $10^{12}\mathrm{\, M_{\odot}}$ into the median SMHM ratio. Histogram bins reveal a continued quenching of the sSFR above this threshold, which has led to lower stellar mass fractions in high mass haloes. This is consistent with the turnover seen in \citet{Behroozi_2019}. Nevertheless, beyond $\sim 10^{13.4}\mathrm{\, M_{\odot}}$, the No-X-ray run still slightly overpredicts the relation. In contrast, the \citet{Behroozi_2019} median remains within the $\rm 16^{th}$ to $\rm 84^{th}$ percentile range in the full Simba-50 run up to the highest halo masses. The activation of X-ray feedback further enhances the dampening of the sSFR for massive haloes, with an increased number of bins exhibiting a sSFR values as low as $10^{-11}\, \mathrm{yr}^{-1}$.

In summary, we find that the inclusion of \textsc{Simba}'s X-ray mode has slightly reduced the SMHM ratio in massive haloes before $z=2$, while the AGN winds and the jets alone have not altered the histograms significantly compared to the No-AGN run. At $z=0$, the suppression of star formation due to \textsc{Simba}'s radiative AGN winds has limited stellar mass growth in haloes ranging from $\sim 10^{11}$ to $10^{12}\mathrm{\, M_{\odot}}$. At higher halo masses, AGN jets produce the turnover of the SMHM ratios at around $10^{12}\mathrm{\, M_{\odot}}$. Adding the X-ray heating mode sees a further quenching of the sSFR at high halo masses and slightly decreased SMHM ratios with respect to the No-X-ray run.

\subsection{AGN feedback effects in individual galaxies at $z \leq$ 2}
\label{sec:AGN feedback z2}
Thus far, we have examined the mass- 
and redshift-dependence of stellar and AGN feedback-induced quenching in both global statistics, as well as at fixed halo mass. In this section, we will further trace the efficiency of \textsc{Simba}'s AGN feedback modes to the properties of individual galaxies and their black holes. 

\subsubsection{sSFR-$M_{\star}$ relation}
\label{sec:sSFR_Mstar}
\begin{figure*}
	% To include a figure from a file named example.*
	% Allowable file formats are eps or ps if compiling using latex
	% or pdf, png, jpg if compiling using pdflatex
	\includegraphics[width=\textwidth]{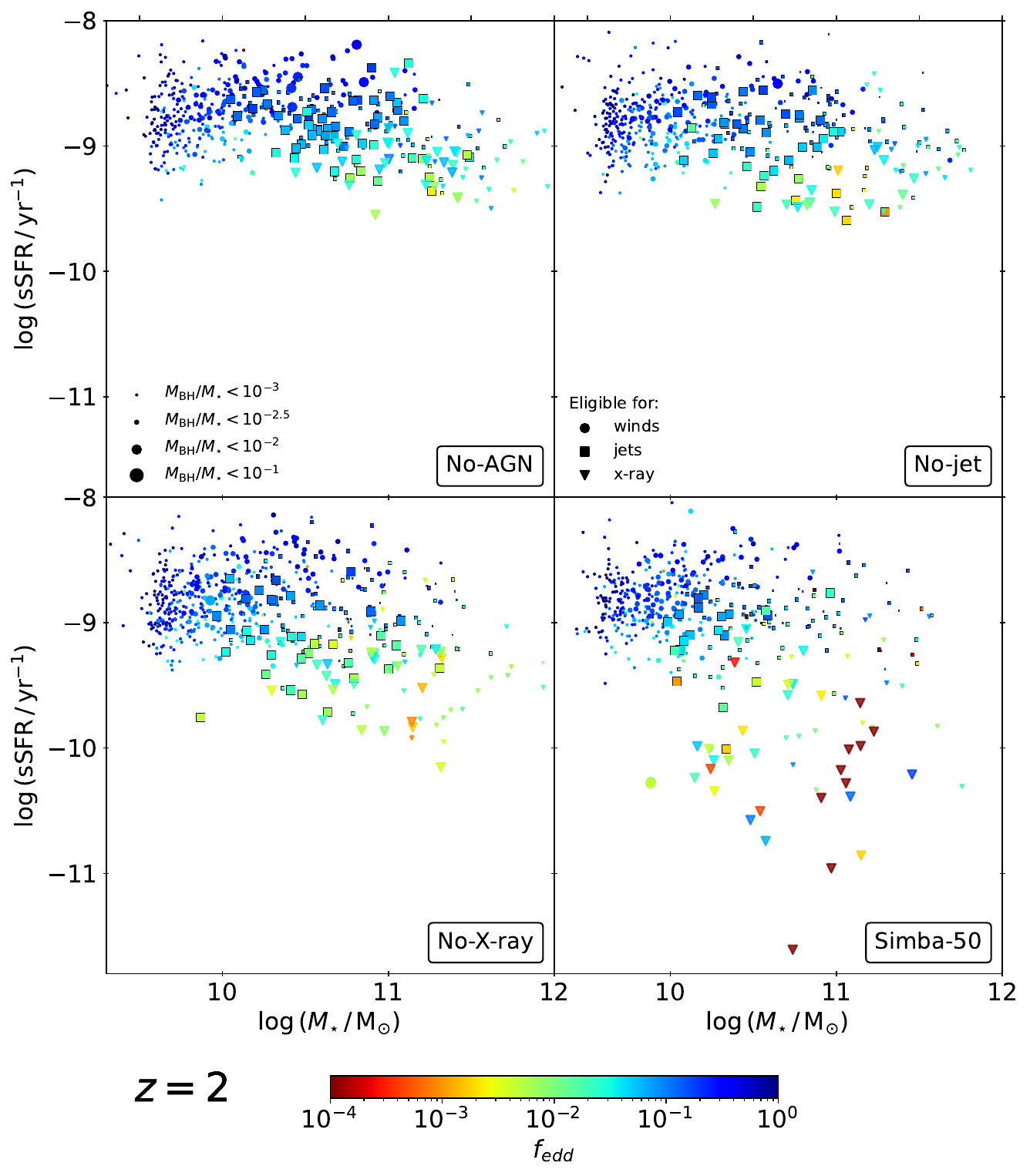}
    \caption{Scatter plots of the sSFR-$M_{\star}$ relation for all central galaxies at $z=2$ in the four runs containing feedback, coloured by Eddington ratio of the AGN. Galaxies eligible for AGN winds are shown in circles, jets in squares (with a black outline as a visual aid), and X-ray feedback in triangles. The marker size indicates the mass of the black hole relative to the galaxy's stellar mass, as indicated in the legend. While some galaxies in the No-X-ray run show slightly reduced sSFR's, only the Simba-50 run contains quenched galaxies at this epoch.}
    \label{fig:SFR_Mstar 2}
\end{figure*}

\begin{figure*}
	% To include a figure from a file named example.*
	% Allowable file formats are eps or ps if compiling using latex
	% or pdf, png, jpg if compiling using pdflatex
	\includegraphics[width=\textwidth]{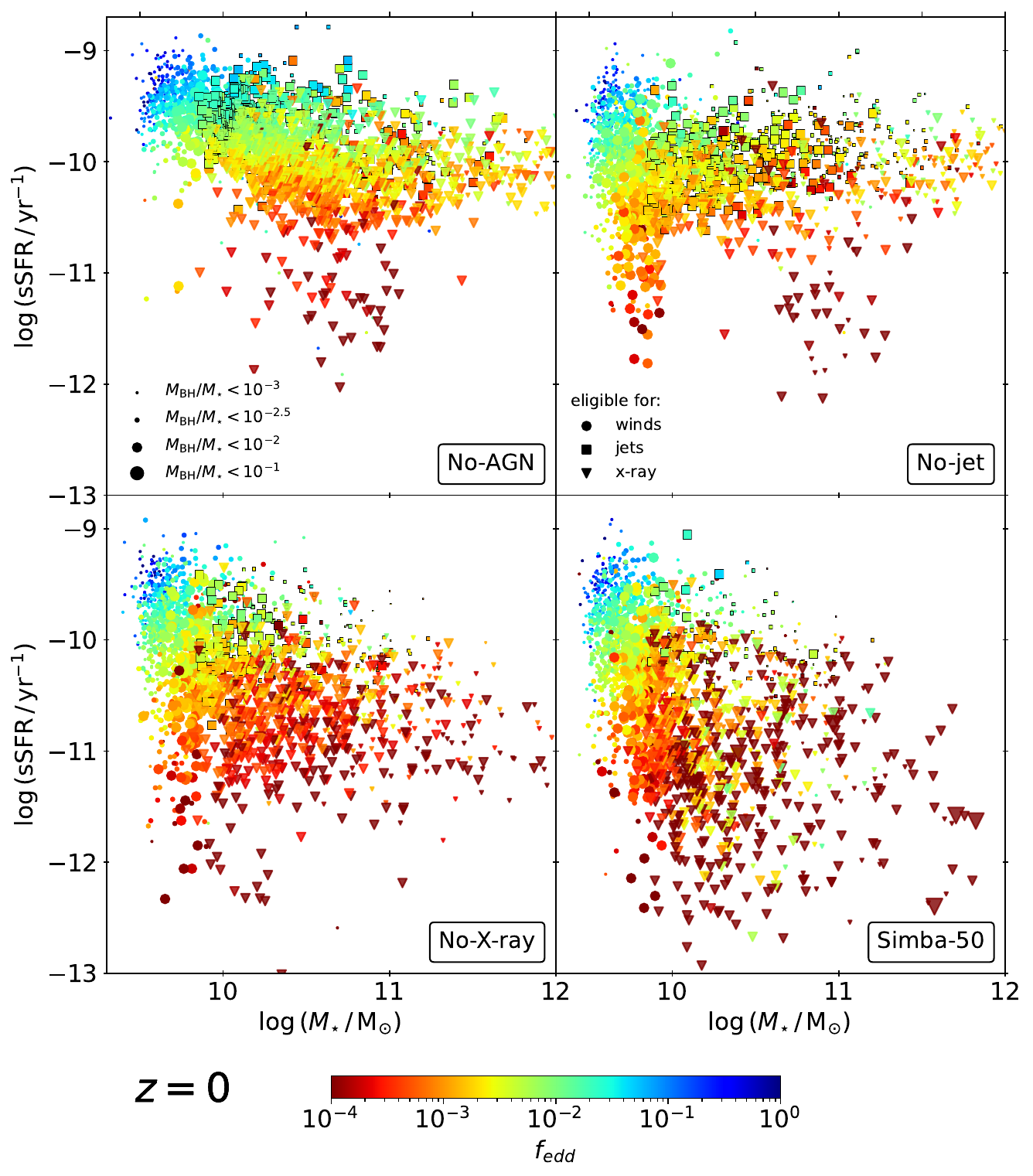}
    \caption{Same layout as Fig. \ref{fig:SFR_Mstar 2}, except at $z=0$. Including AGN winds reduces sSFR's for intermediate-mass galaxies below $10^{10.5} \, \rm M_{\odot}$. Jets quench galaxies above $10^{10} \, \rm M_{\odot}$ and introduce a correlation between the sSFR and the accretion rate. This correlation is broken by the X-ray mode in the Simba-50 run, where although high mass galaxies are further quenched, Eddington ratios are more variable.}
    \label{fig:SFR_Mstar 0}
\end{figure*}

In Fig. \ref{fig:SFR_Mstar 2} ($z=2$) and \ref{fig:SFR_Mstar 0} ($z=0$), we consider the scatter of the sSFR-$M_{\star}$ relation for all central galaxies with a black hole particle in the No-AGN, No-jet, No-X-ray, and Simba-50 runs. Galaxies have been categorised according to their eligibility for each of the three AGN feedback modes, regardless of whether a particular mode is included in the simulation run. Galaxies are represented by circular, square and triangular markers if they satisfy the eligibility criteria for AGN radiative winds, AGN jets, and X-ray feedback, respectively (see \S~\ref{sec:simulations}). As we consider only galaxies with a central black hole, this figure excludes the smallest galaxies below $10^{9.5}\mathrm{\, M_{\odot}}$.

As the mass of the black hole directly scales the strength of the AGN wind mode, as well as determines the sphere of influence of the black hole, we further indicate the ratio of the black-hole-to-stellar mass ratio in the marker size.
We distinguish between $M_\mathrm{BH}/M_{\star} < 10^{-3}$, $10^{-3} < M_\mathrm{BH}/M_{\star} < 10^{-2.5}$, $10^{-2.5} < M_\mathrm{BH}/M_{\star} < 10^{-2}$,  and $10^{-2} < M_\mathrm{BH}/M_{\star} < 10^{-1}$. Lastly, we colour-code markers by the Eddington ratio of the black hole. We note, that all square markers, signifying eligibility for jet feedback (but not X-ray feedback), have been given a black outline as a visual aid.

At $z=2$, galaxies in the No-AGN run exhibit sSFR's between $10^{-9.5}$ and $10^{-8}\, \mathrm{yr}^{-1}$, with no clear trend across all stellar masses. This finding aligns with our results from earlier sections, where we established that \textsc{Simba}'s stellar feedback has minimal impact on high-mass galaxies. Broadly, the type of feedback a galaxy is eligible for depends on stellar mass. Galaxies that would be eligible solely for radiative winds, are concentrated among lower stellar masses, typically between $10^{9.5}$ and $10^{10}\mathrm{\, M_{\odot}}$. Between roughly $10^{10}$ and $10^{11}\mathrm{\, M_{\odot}}$, most galaxies are eligible for jet feedback. At even higher stellar masses, an increasing number of galaxies would qualify for X-ray feedback as well. This division remains in subsequent plots and is a consequence of the co-evolution  of galaxies and black holes. As none of these prescriptions are included in the No-AGN run, we do not expect any effect due to AGN feedback. However, while the accretion rate is high for most AGN in this run, a few galaxies at the higher mass end exhibit lowered values for $f_\mathrm{edd}$. These coincide with slightly lower sSFR's, which illustrates that since both accretion and star formation draws from the gas reservoir of the host galaxy, these processes are intimately connected, even in the absence of AGN feedback prescriptions.

The No-jet run follows a similar pattern to the No-AGN run, in agreement with our findings from previous sections that at $z=2$, radiative winds have not yet significantly affected galaxies. A few galaxies exhibit low accretion rates, with slightly lowered sSFR's, but they generally remain in the $10^{-9.5}-10^{-8}\, \mathrm{yr}^{-1}$ range, with no qualitative difference from the No-AGN run. 
In the No-X-ray run, this effect becomes more pronounced and the sSFR in some galaxies with jet and X-ray eligibility is suppressed to around $10^{-10}\, \mathrm{yr}^{-1}$, a commonly used threshold for a galaxy to still be classified as star-forming. This affects most galaxies with relatively large $M_\mathrm{BH}/M_{\star}$ and accretion rates below $10^{-1}$. In the Simba-50 run, the majority of such galaxies exhibit quenched sSFR's between $10^{-10}$ and $10^{-11}\, \mathrm{yr}^{-1}$. Almost all of these affected galaxies are eligible for X-ray feedback, explaining why this feature is only seen in this full physics run, when X-ray feedback is actually activated. These galaxies are the origin of the gap between the SFRD evolution, in the No-X-ray and Simba-50 runs, shown in Fig. \ref{fig:SFRD}. However, it is surprising that only a few tens of quenched galaxies in this run make up the difference of 0.2\,dex in the overall star formation at $z \sim 2$. We investigate this feature further in the following section.

At $z=0$, the sSFR in the entire galaxy population in the No-AGN run is at least 1 dex lower relative to the $z=2$ snapshot. We note that the quenched galaxies with sSFR between $10^{-11}$ and $10^{-12} \mathrm{yr^{-1}}$ and stellar masses between $10^{10}$ and $10^{11}\mathrm{\, M_{\odot}}$, match the population with high stellar mass fractions in Fig. \ref{fig:heatmapz0}, which we identified as probable halo finder artefacts. We thus exclude these galaxies, and the analogous population in the No-jet plot, from further analysis.

In the No-jet run, the AGN winds quench the sSFR in galaxies with a relatively large black hole and stellar masses between $10^{9.5}$ and $10^{10.5}\mathrm{\, M_{\odot}}$. Additionally, there is a trend toward generally lower stellar masses, while in the No-AGN run, most galaxies accumulate at intermediate stellar masses. This suggests that feedback from AGN winds has slowed the growth of stellar masses for smaller galaxies. Larger galaxies appear unaffected by the inclusion of AGN winds, in agreement with our earlier findings. 

Activating the jets in the No-X-ray run results in the quenching of most galaxies with stellar mass $> 10^{10}\mathrm{\, M_{\odot}}$, which again demonstrates the powerful impact of jet feedback. This creates a clear downward slope, where above $10^{11}\mathrm{\, M_{\odot}}$, there are almost no galaxies with $\rm sSFR > 10^{-10}\mathrm{yr^{-1}}$, which dominate the No-jet run. Low-mass galaxies continue to exhibit a wide range of sSFR's between $10^{-9}\mathrm{yr^{-1}}$ and $10^{-12}\mathrm{yr^{-1}}$. We further note a clear correlation between $f_\mathrm{edd}$ and the sSFR, where the most severely quenched galaxies also exhibit the lowest accretion rates. Most of these quenched galaxies would also be eligible for X-ray feedback, if it was included in this run, meaning these galaxies exhibit low gas fractions. This is in contrast with the No-jet run, which showed much more galaxies above $10^{10}\mathrm{\, M_{\odot}}$ as eligible for jets, meaning higher gas fractions.   

With the inclusion of the X-ray feedback mode in the Simba-50 run, the sSFR's are further suppressed in many galaxies, with a significant number of them exhibiting values below $10^{-11.5}\mathrm{yr^{-1}}$. There is a larger fraction of galaxies in the $10^{9.5}-10^{10}\mathrm{\, M_{\odot}}$ range, indicating that the impact of X-ray heating has been sufficient to slow down the growth of stellar mass in individual galaxies. While the No-X-ray run showed a clear correlation between the Eddington ratio and the sSFR across all galaxies, we find that in the full physics run, galaxies affected by X-ray feedback exhibit a much larger scatter in $f_\mathrm{edd}$. Two galaxies of similar stellar mass and similarly quenched sSFR's, might differ up to 2~dex in their Eddington fraction. This is surprising, since both accretion and star formation draw from the gas reservoir of the galaxy. It is likely a consequence of the central energy input from the X-ray mode, which increases the relative contribution of Bondi accretion from hot gas to the overall accretion rate and suppresses local star formation. We explore this connection further in \S~\ref{sec:xray_BH}. Lastly, we note that more galaxies in this run contain galaxies with particularly large black holes for their stellar masses, suggesting that the activation of the X-ray mode in some cases allows for more efficient black hole growth. This occurs because once a black hole is growing via Bondi accretion, its accretion rate scales as $M_\mathrm{BH}^2$, which for massive black holes can result in rapid growth.

\subsection{Impact of X-ray feedback on highly-star forming galaxies and black-hole growth}
\label{sec:xray}

\subsubsection{Cosmic star formation rate density in star-forming and quiescent galaxies}
\label{sec:xray_SFR}

\begin{figure*}
	% To include a figure from a file named example.*
	% Allowable file formats are eps or ps if compiling using latex
	% or pdf, png, jpg if compiling using pdflatex
	\includegraphics[width=\textwidth]{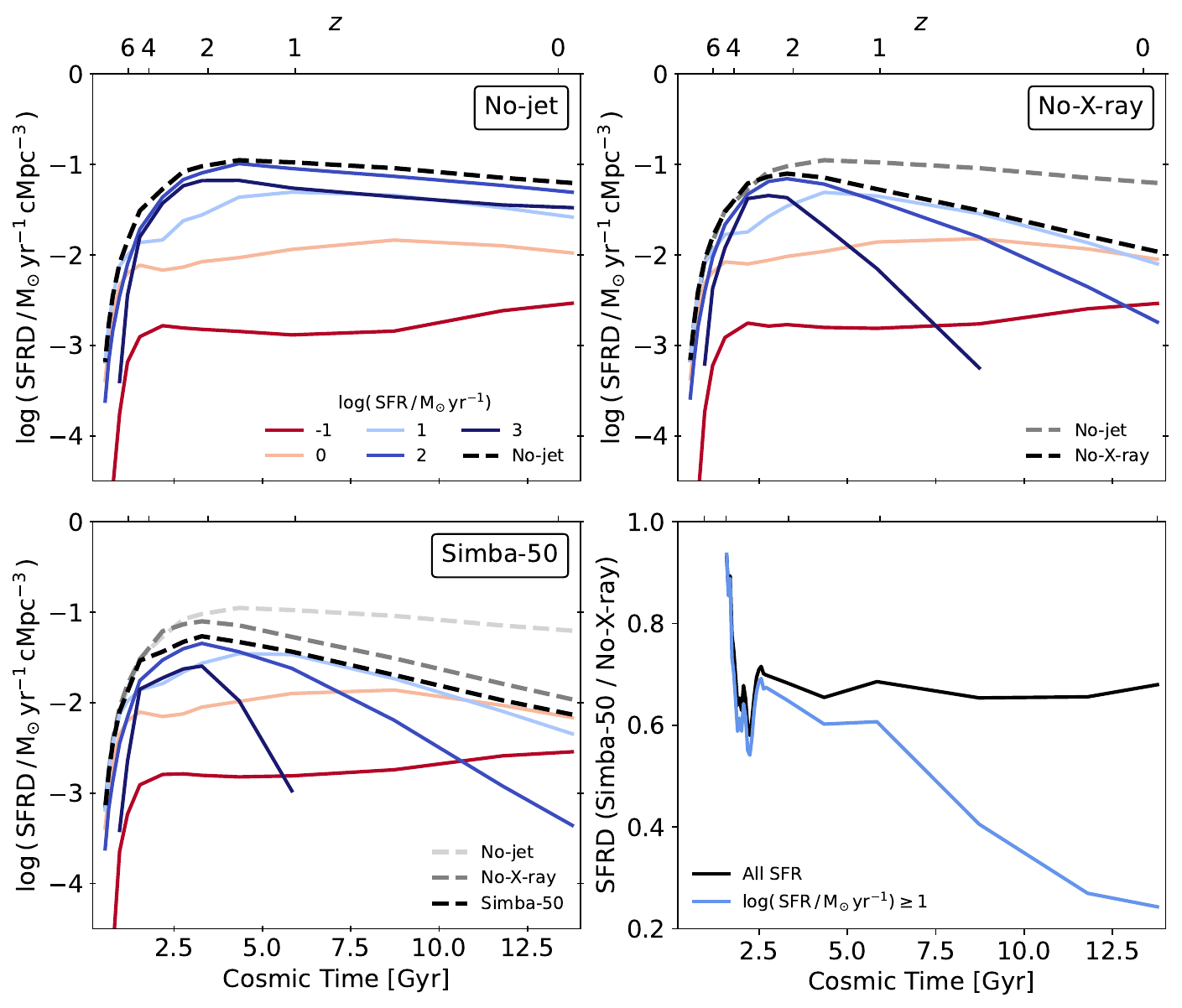}
    \caption{\textit{Top panels and bottom left panel:} SFRD in galaxies in bins of SFR between $\rm log(SFR/M_{\odot} yr^{-1})$ = -1 and 3 (see legend). Different panels show the No-jet, No-X-ray and Simba-50 run, with the total SFRD from all galaxies in the indicated run (black dashed line). As a visual reference, we further include the total SFRD from the previous panel (grey dashed lines). The highest SFR bins suffer the most quenching by jets and X-ray heating. For X-ray feedback, this suppression of the SFRD sets in before $z=2$, while the jets are only efficient after $z=2$. \textit{Bottom right panel:} Ratio between the SFRD in the No-X-ray and the Simba-50 run from galaxies with $\mathrm{SFR > 10 \, M}_{\odot} \, \mathrm{yr}^{-1}$ (blue) and all galaxies (black). Before $z=2$, the ratio between the SFRD from highly star-forming galaxies in the No-X-ray and Simba-50 runs closely traces the ratio of the SFRD from all galaxies. This indicates that the quenching of these galaxies due to X-ray heating is responsible for the early discrepancy between these runs.}
    \label{fig:SFRDSFR}
\end{figure*}

In prior sections, we have shown repeatedly that the X-ray heating mode in \textsc{Simba} has a non-negligible effect on star formation even before $z = 2$. The star formation rate density in Fig. \ref{fig:SFRD} showed a discrepancy of around 0.2\,dex between the curves for the No-X-ray and Simba-50 runs. The SMHM ratio at $z=2$ (Fig. \ref{fig:heatmapz2}) exhibited lower stellar mass fractions at high halo masses, as well as some bins with suppressed specific star formation rates for the Simba-50 run, suggesting that this early quenching of a few galaxies is enough to alter the overall distribution of stellar mass in the simulation. In \S~\ref{sec:sSFR_Mstar}, we identified a small number of individual galaxies which presented suppressed sSFR's at $z=2$. In this last section, we investigate why these galaxies are sufficient to lower the overall SFRD before $z=2$. 

The quenching of these galaxies would be fractionally substantial if the X-ray mode primarily suppressed star formation in previously highly star-forming galaxies.
To assess the overall contribution to the SFRD from galaxies at different SFR, we bin all galaxies into five equally spaced logarithmic bins, which are centred on $\mathrm{10^{-1}}, 1, 10, 10^{2}$, and $\mathrm{10^{3} \, M}_{\odot} \, \mathrm{yr}^{-1}$ and extend 0.5\,dex above and below these values. In Fig. \ref{fig:SFRDSFR}, we display the SFRD computed for galaxies in each of these bins (see colour-code in legend), alongside the total SFRD (black dashed line) in the No-jet, No-X-ray and Simba-50 runs (top panels and bottom left panel). To guide the comparison, we further include the total SFRD from the runs in preceding panels (grey dashed lines).

In the No-jet run, the SFRD contribution from all SFR bins remains relatively constant, with particularly high contributions from the highest SFR bins, specifically $10 \rm \, M_{\odot} \, \mathrm{yr}^{-1}$ to $\mathrm{10^{3} \, M_{\odot} yr^{-1}}$ (blue lines). However, upon activating jets in the No-X-ray run, there is already a noticeable decrease in the SFRD contributions from galaxies in these SFR bins when compared to the No-jet run. As a result, the $\mathrm{10 \, M}_{\odot} \mathrm{yr}^{-1}$ bin begins to set the shape of the SFRD after $z \sim 1$. While the suppression of the $10^{2}$ and $\mathrm{10^{3} \, M_{\odot}} \mathrm{yr}^{-1}$ bins begins just before $z\sim2$, the overall SFRD is not lowered significantly until after $z \sim 2$. Instead, an increased SFRD contribution from the $\mathrm{10 \, M}_{\odot} \mathrm{yr}^{-1}$ bin before $z\sim 2$ compensates for the early suppression.

In contrast, the Simba-50 run shows that the $\mathrm{10^{2} \, M}_{\odot} \mathrm{yr}^{-1}$ and $\mathrm{10^{3} \, M}_{\odot} \mathrm{yr}^{-1}$ bins are quenched by 0.2--0.3~dex from $z \sim 4$, and to a lesser extent, so is the $\mathrm{10 \, M}_{\odot} \mathrm{yr}^{-1}$ bin. Lower SFR bins appear largely unchanged. This suggests that the quenching of the overall SFRD before $z=2$ is indeed primarily due to the reduced contribution from highly star-forming galaxies under X-ray heating.

We further verify that the quenching of formerly highly star-forming galaxies is quantitatively responsible for the difference in SFRD at $z\sim2-4$ between the No-X-ray and Simba-50 runs. In the bottom right panel of Fig. \ref{fig:SFRDSFR}, we show the ratio of the SFRD from all galaxies (black) in both runs and the SFRD ratio of the galaxies with $\mathrm{SFR > 10 \, M}_{\odot} \mathrm{yr}^{-1}$ (blue). Up to just before $z=2$, the ratio of the SFRD from all galaxies in the entire Simba-50 run with respect to the No-X-ray run is mirrored by the SFRD ratio computed from only the highly star-forming galaxies. After $z=2$, the SFRD ratio from these galaxies decreases to 0.2--0.3, meaning residual quenching of lower-SFR galaxies due to X-ray heating only contributes significantly to the gap between the No-X-ray and Simba-50 SFRD's at late times. Thus we conclude that the gap between $z=2-4$ across the No-X-ray and Simba-50 runs in Fig. \ref{fig:SFRD} can be fully explained by the quenching of a small number of galaxies that were initially highly star-forming. 

\subsubsection{Distribution of sSFR for galaxies on the $f_{\rm edd}$-$M_{\rm BH}$ relation}
\label{sec:xray_BH}
\begin{figure*}
	\includegraphics[width=0.98\textwidth]{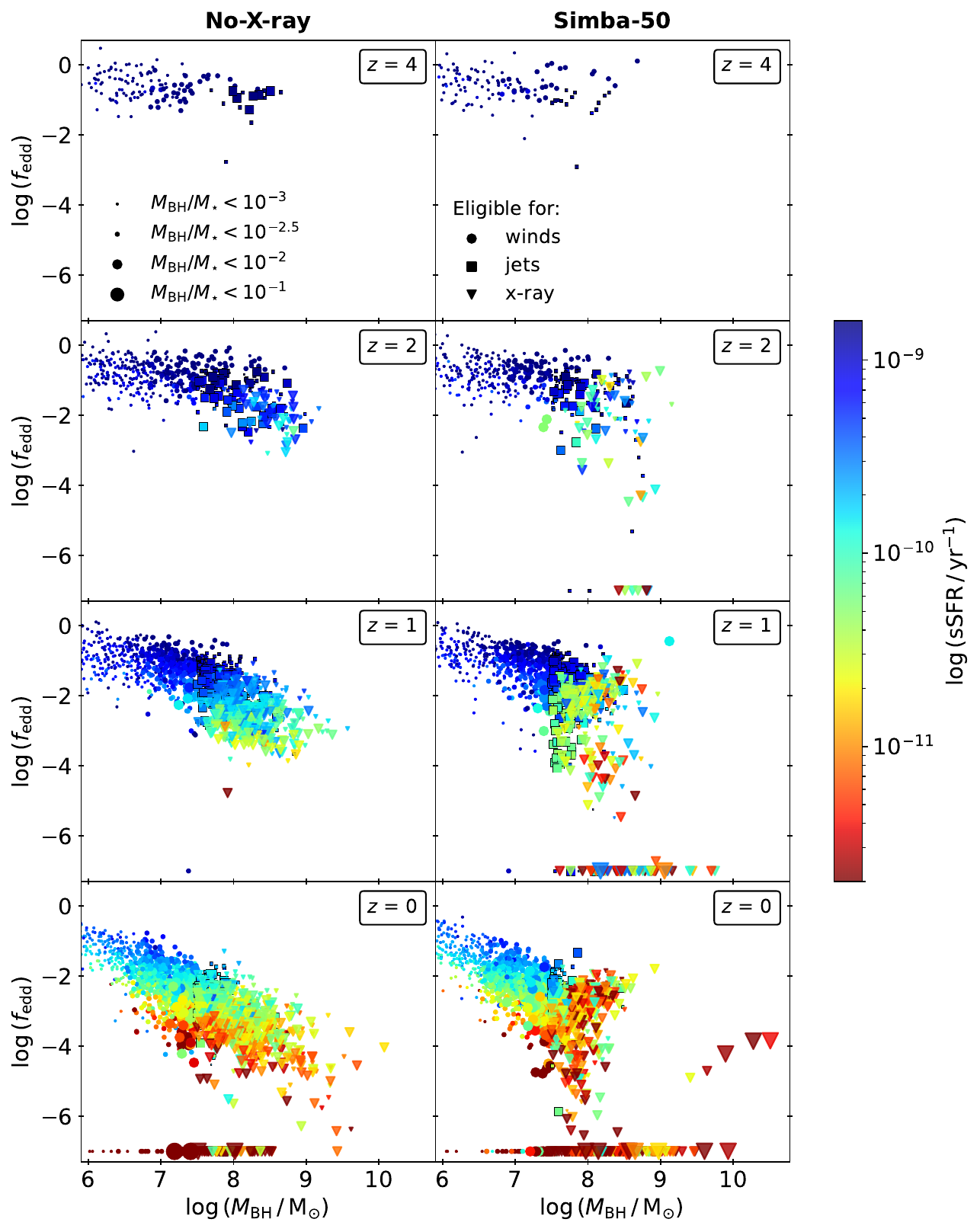}
    \caption{Scatter plots of the $f_{\rm edd}$-$M_{\rm BH}$ relation for all central galaxies in the No-X-ray and Simba-50 runs between $z=4-0$, coloured by sSFR. As in Fig. \ref{fig:SFR_Mstar 2}, galaxies eligible for AGN winds are shown in circles, jets in squares (with a black outline as a visual aid) and X-ray feedback in triangles, with the marker size indicating the size of the black hole. While the No-X-ray run shows a clear anti-correlation between the $f_{\rm edd}$ and $M_{\rm BH}$, the inclusion of X-ray feedback in the Simba-50 run introduces a break above $M_\mathrm{BH} = 10^{7.5}\,\rm M_{\odot}$.}
    \label{fig:fedd_MBH}

\end{figure*}

In this section, we aim to explain why the $f_\mathrm{edd}$-sSFR correlation breaks down upon the activation of the X-ray mode, as seen in Fig. \ref{fig:SFR_Mstar 0}. Additionally, we further examine the relationship of the galaxy sSFR's to the accretion rate and the black hole mass.
In Fig. \ref{fig:fedd_MBH}, we show scatter plots of the $f_{\rm edd}$-$M_{\rm BH}$ relation for all central galaxies in the No-X-ray and Simba-50 runs at $z=4$, 2, 1, and 0. Markers are coloured by the galaxy's sSFR. Similar to Fig. \ref{fig:SFR_Mstar 2}, we represent galaxies eligible for AGN winds with circular markers, those eligible for jets with square markers (outlined in black as a visual aid), and galaxies eligible for X-ray feedback with triangular markers. The size of each marker corresponds to the ratio of black hole mass to stellar mass of the host galaxy. 

At $z=4$, all galaxies within both runs exhibit high sSFR's with no discernible signs of quenching. The majority of these galaxies are primarily influenced by AGN winds, given that most of their black holes are still below the $M_\mathrm{BH} = 10^{7.5}\, \rm M_{\odot}$ threshold. However, a few larger galaxies beyond this threshold feature sufficiently low accretion rates to be eligible for jet feedback. We note a slight tendency toward higher $M_{\rm BH}/M_{\star}$ ratios in the No-X-ray run. In the Simba-50 run, temporary activity of the X-ray mode may have caused slightly lower stellar masses within these galaxies.

At $z=2$, the runs show more pronounced differences. Both runs now include galaxies eligible for X-ray feedback, which contributes to the emergence of quenched galaxies in the Simba-50 run, consistent with Fig. \ref{fig:SFR_Mstar 2}. These galaxies further exhibit accretion rates below $\rm log (\mathit{f}_{edd}) = -3.5$, which similarly do not occur in the No-X-ray run. Some galaxies with low Eddington fractions are only eligible for jets. They are characterised by relatively small black hole masses and exhibit high sSFR's, indicating that they are not quenched. These low accretion rates are likely a result of the small spheres of influence of the black holes and the reduced gas concentrations in the galactic centre due to temporary X-ray feedback.

At $z=1$, quenched galaxies are also observed in the No-X-ray run. Galaxies display a positive correlation between the sSFR and the $f_{\rm edd}$, which in turn shows a clear anti-correlation with $M_{\rm BH}$. In contrast, galaxies eligible for X-ray feedback in the Simba-50 run show more variable accretion rates and sSFR's. Black holes with zero accretion rates have been assigned $\rm log (\mathit{f}_{edd}) = -6.5$. Additionally, a population of galaxies with $M_{\rm BH} \sim 7.5$, which are only eligible for jets, not X-ray, exhibits significantly decreased accretion rates below $\rm log (\mathit{f}_{edd}) = -2$. This population is notably absent in the No-X-ray run. It is likely that these galaxies were affected by X-ray feedback, causing rapid quenching of their accretion rates, and a subsequent re-accretion of some gas led to higher gas fractions, deactivating the X-ray mode. A further investigation showed that gas fractions in these galaxies indeed hover around 0.2, the threshold for X-ray feedback activation.

At $z=0$, the anti-correlation between $f_{\rm edd}$ and $M_{\rm BH}$, along with a correlation with sSFR, persists in the No-X-ray run. In the Simba-50 run, the anti-correlation holds until $M_\mathrm{BH} = 10^{7.5}\, \rm M_{\odot}$, above which the X-ray mode causes considerable variations in Eddington ratios and sSFR's of up to 2.5~dex and 1.5~dex, respectively.

\citet{Thomas2019BlackSimba} found the same break in the $f_{\rm edd}$- $M_{\rm BH}$ anti-correlation in the Simba-100 run. They ascribe this to variable accretion rates due to the Bondi accretion mode for hot gas, which is inherently less stable than the tidal accretion mode for cold gas. The sudden input of thermal energy close to the black hole could transform a significant amount of ISM-gas into non-ISM gas, which would both quench the star formation rate and increase the contribution from Bondi accretion. We explain this reasoning in more detail in \S~\ref{sec:disc_xray}.

\section{Discussion}
\label{sec:discussion}
\subsection{Feedback effects on the cosmic star formation history in \textsc{Simba}: A cohesive picture}
In previous sections, we examined various observables related to star formation and stellar masses and subsequently contrasted their evolution in different \textsc{Simba} simulation variants with isolated feedback prescriptions. In this section, we unify these results into a cohesive picture of how the cosmic star formation history is affected by stellar and AGN feedback in the full-physics \textsc{Simba} simulations. We proceed by explaining how each feedback mode acting in individual galaxies at specific redshifts gives rise to key features in global observables, such as the SFRD and the GSMF.

\subsubsection{Stellar feedback}
\label{sec:disc_stellar}
From the beginning of star formation at early epochs, galaxies are subject to the effects of \textsc{Simba}'s stellar feedback mode. This mainly affects further star formation due the ejection of particles from mass-loaded two-phase winds, resulting in galaxy outflows, which significantly reduce the availability of cold gas for star formation and thus curtail the star-forming activity in these galaxies. Due to the inverse scaling of the mass-loading factor with stellar mass, according to a power law with break at $\mathrm{M_{0}=5.2 \times 10^{9}\, M_{\odot}}$ (see Eq. \ref{eq:massloading}), as well as the mass scaling of the counteracting gravitational potential, stellar feedback is more effective in low-mass haloes, which are more prevalent at high redshifts. 

As a result of stellar feedback, the median nSFR in halo populations with low total masses ($M_\mathrm{H}<10^{12.5}\rm \, M_{\odot}$) is lowered in all runs relative to the No-feedback run (see Fig. \ref{fig:halobins}). This is entirely responsible for the high-redshift suppression of the SFRD by 0.8 dex until $z\approx 4$, after which AGN feedback starts suppressing and even quenching star formation in more massive haloes (see Fig. \ref{fig:SFRD}). After $z=4$, stellar feedback contributes a continued, albeit decreasing, suppression of the SFRD until $z=0$. This slows stellar mass growth in low-mass systems and, consequently, the GSMF (Fig. \ref{fig:GSMF}) is reduced by 1~dex at all masses for $z = 4$ and below $M_\mathrm{\star} \sim 10^{10}\rm \, M_{\odot}$ at $z=1$ and $0$.

\subsubsection{AGN winds}
\label{sec:disc_winds}
As stellar masses gradually increase, albeit at a slowed pace due to the influence of stellar feedback, galaxies eventually reach stellar masses of $M_{\star} \gtrsim 10^{9.5}\mathrm{\, M_{\odot}}$. At this point the first black holes are seeded, with seed masses of $10^{4}\mathrm{\, M_{\odot}}h^{-1}$. These newly formed black holes exhibit high accretion rates, and AGN winds are active immediately. However, the wind velocities scale with the black hole mass and are thus low initially. For $M_\mathrm{BH} = 10^{4}\mathrm{\, M_{\odot}}h^{-1}$, the wind velocity is approximately $\approx 170 \rm \, km \, s^{-1}$ (see Eq.~\ref{eq:winds}). This may be insufficient to exert significant influence on the star forming activity. During this early phase, galaxies remain gas-rich, fuelling high accretion and star formation rates. At around $z \sim 2$, Fig. \ref{fig:SFR_Mstar 2} shows only a few instances of slightly lowered accretion rates due to the influence of AGN winds. 

At later cosmic epochs, black holes have grown more massive and accelerated AGN wind particles can achieve velocities of up to $\sim 1200 \rm km \, s^{-1}$ for black hole masses of $10^{10}\mathrm{\, M_{\odot}}$. This mode is relevant for quenching intermediate-mass systems, where black hole masses are not yet large enough for jets to have taken over as the dominant quenching mode. As shown in Fig. \ref{fig:SFR_Mstar 0}, galaxies at $z = 0$ with $M_{\star} \sim 10^{9.5}-10^{10}\mathrm{\, M_{\odot}}$ and relatively large black holes at low accretion rates ($f_\mathrm{edd} < 0.01$) exhibit sSFR's below $10^{-11} \mathrm{yr^{-1}}$. As a result of the slowed stellar mass growth, galaxies accumulate at this mass range at the expense of more massive systems. This translates into the minor peak in the GSMF for galaxies with stellar masses around $10^{9.5}-10^{10}\mathrm{\, M_{\odot}}$ observed in Fig.~\ref{fig:GSMF}. 

The quenching of intermediate-mass galaxies further manifests in the SMHM ratio at $z=0$, as shown in Fig. \ref{fig:heatmapz0}. In this case, stellar mass fractions are reduced within haloes in the mass range of $M_\mathrm{H} \sim 10^{11}-10^{12}\mathrm{\, M_{\odot}}$. The suppression of the sSFR in individual galaxies has a cumulative effect on the median nSFR after $z \sim 1$ in Fig. \ref{fig:halobins}, where haloes with $M_\mathrm{H} \sim 10^{12}\rm \, M_{\odot}$ exhibit a noticeable drop of 0.5~dex in the No-jet run. At higher masses, the escape velocity exceeds the maximum achievable wind speed (1200$\mathrm{~km} \mathrm{~s}^{-1}$), which renders the wind mode insufficient to quench the nSFR.

\subsubsection{Jet feedback}
\label{sec:disc_jets}
The first galaxies eligible for the jet mode appear around $z=4$, as shown in Fig. \ref{fig:fedd_MBH}. However, their sSFR's remain relatively high, at around $10^{-9} \mathrm{yr^{-1}}$. Fig. \ref{fig:SFR_Mstar 2} shows that upon reaching $z = 2$ in the No-X-ray run, individual high-mass galaxies with jets and low accretion rates ($f_\mathrm{edd} < 10^{-2.2}$) exhibit sSFR's below $10^{-9.5}\mathrm{yr^{-1}}$, which are not present in the No-jet run. This trend can also be seen in the SFRD (Fig. \ref{fig:SFRD} and \ref{fig:SFRDSFR}). It coincides with a reduced number of galaxies at high stellar masses with respect to the No-jet run (see Fig. \ref{fig:SFR_Mstar 2} and Fig. \ref{fig:GSMF}). However, jets are in general not efficient at fully quenching galaxies at these epochs, except in special cases that have specific environmental conditions~\citep{Szpila2024TheSimulation}.

After $z=2$, however, jets become the dominant quenching mechanism in \textsc{Simba}. By $z=0$, they have significantly altered the slope of the sSFR-$M_{\star}$ relation in the No-X-ray run, particularly impacting galaxies with $M_{\star} \gtrsim 10^{10} \,\mathrm{M}_{\odot}$ (Fig.~\ref{fig:SFR_Mstar 0}). This translates into a reduction of the median nSFR by $1-1.5\,$dex for halo populations with masses greater than $10^{12.5} \, \mathrm{M}_{\odot}$, as well as a decrease of the overall SFRD by 1\,dex. Stellar mass growth is slowed drastically, as a result, and the jets crucially introduce a downward bend into SMHM ratio at $M_\mathrm{H} \sim 10^{12} \, \mathrm{M}_{\odot}$ and reduce the GSMF by up to 1~dex at $M_{\star} \gtrsim 10^{11} \, \mathrm{M}_{\odot}$.

We further find a clear correlation of $f_\mathrm{edd}$ and sSFR in the No-X-ray run. Both star formation and accretion processes draw from the galaxy's gas reservoir and they thus correlate with the amount of available gas. Gas is more easily heated and expelled from the galaxy for higher outflow velocities. Since the speed of AGN jets scales inversely with the logarithmic accretion rate down to $f_\mathrm{edd} = 0.02$ (Eq.~\ref{eq:jet}), then a lower accretion rate is associated with lower sSFR.  

The efficiency of the jet mode at late times is connected to the detailed mechanism powering this AGN feedback mode, and the redshift evolution of the properties of the circumgalactic medium (CGM). 
The particles that are ejected by the AGN jets are heated to the virial temperature of the halo and subsequently deposited in the CGM outside the galaxy, at distances of around $50 \, \rm kpc$. Ejected particles are initially decoupled, meaning that they undergo no hydrodynamic interactions with the surrounding gas. After $10^{-4}$ Hubble times, particles recouple and deposit their high thermal energy in the CGM. 
Before $z \sim 2$, the CGM is characterised by high densities. The heating introduced by AGN feedback thus tends to dissipate quickly due to elevated cooling rates. As a result, AGN jets are not effective in high-redshift environments. At lower redshifts, the CGM is more diffuse and retains most of the deposited heat. \citet{Appleby2021TheSimba} showed that a low redshifts, the CGM of No-X-ray and Simba-50 galaxies with $M_{\star} \gtrsim 10^{10} \, \mathrm{M}_{\odot}$ is primarily composed of hot gas, with temperatures $>0.5\,T_{\rm vir}$. This restricts the (re-)accretion of CGM gas onto the galaxy, creating a scenario where galaxies effectively starve due to low gas availability, resulting in a slow quenching of the star formation rate and the accretion rate over dynamical timescales \citep[see slow quenching described in][]{Montero2019MergersSimulation}. Importantly, this form of AGN feedback is mostly preventative, and does not significantly affect the local ISM within galaxies. Instead, only a small fraction of the ejected particles, approximately a $1/50$ of the gas mass, is carried out by the feedback process. 

Jets affect particularly high-mass haloes, as they typically contain larger and more mature black holes with low accretion rates, which are required for the jets to initially activate and then reach velocities high enough for heated gas particles to leave the galaxy. Additionally, the virial temperature scales with halo mass, resulting in larger amounts of thermal energy being carried by the jets. The vast majority of jet energy is kinetic, however, as the virial velocity of massive halos is $\sim 10^3$\,K while the jet velocity is $\sim 10^4$\,K.

We note that previous work with the \textsc{Simba} suite of simulations has also identified $z=2$ as a critical epoch. \cite{Sorini2022} showed that before this redshift, stellar feedback is the dominant mechanism shaping the distribution of baryons in the IGM, and within low-mass haloes and their CGM. After $z=2$, the AGN-jet feedback mode exhibits the strongest impact on the baryonic content of high-mass haloes and their CGM, as well as the content and thermal state of the IGM \citep[see also][]{Christiansen_2020, Khrykin_2024}. Our works therefore shows that the relative importance of different feedback modes on star formation and baryon distribution are closely related. The underlying reason is most likely connected to the smaller number of galaxies eligible for AGN-jet feedback before $z=2$ (Fig.~\ref{fig:fedd_MBH}).

\subsubsection{X-ray heating}
\label{sec:disc_xray}
The inclusion of X-ray heating is crucial for producing quenched galaxies before $z = 2$. This is evident in Fig \ref{fig:SFR_Mstar 2}, where a few tens of galaxies in the full Simba-50 run exhibit sSFR's below $10^{-10}\mathrm{yr^{-1}}$, which are notably absent in the No-X-ray run. These galaxies also contain relatively large black holes for their stellar masses. Fig. \ref{fig:SFRDSFR} has confirmed that the quenching of this small number of formerly highly star-forming galaxies (SFR $> 10 \,\mathrm{M_{\odot} \,yr^{-1}}$) is responsible for the diverging SFRD at $z\sim 4$ in the full Simba-50 run in Fig. \ref{fig:SFRD}. Signs of early quenching resulting from the activation of the X-ray mode also manifest in the median nSFR for haloes around $10^{13} \rm \, M_{\odot}$ and a few bins with low sSFR's in the SMHM ratio histogram in Fig. \ref{fig:heatmapz2}. Notably, since these effects are limited to only a handful of galaxies, they have exerted a relatively minor influence on the overall stellar mass distribution at $z=2$. Consequently, no significant differences in the GSMF and only a slight trend toward lower SMHM ratios are evident at this epoch.

After $z = 2$, when the jets can efficiently quench galaxies, \textsc{Simba}'s X-ray heating mode plays a minor role. It remains responsible for suppressing residual star formation in high-mass galaxies, as seen in the sSFR's of individual galaxies with stellar masses exceeding $10^{10} \mathrm{\, M_{\odot}}$ in Fig. \ref{fig:SFR_Mstar 0} and the quenching $\rm \sim 0.5~dex$ of the median nSFR of haloes with $M_\mathrm{H}>10^{12.5} \mathrm{\, M_{\odot}}$. This further results in a continued 0.2~dex suppression of the overall SFRD after $z\sim 2$. The resulting stellar mass growth is only affected at the highest masses, as X-ray feedback reduces the number of galaxies with $M_{\star} = 10^{11}-10^{12} \mathrm{\, M_{\odot}}$ in the GSMF and results in a decline in SMHM ratios for haloes with masses exceeding $10^{13} \mathrm{\, M_{\odot}}$. Lastly, we note that the X-ray mode breaks the anti-correlation between black hole mass and accretion rate (Fig. \ref{fig:fedd_MBH}), as well as the correlation of sSFR and accretion rate (Fig. \ref{fig:SFR_Mstar 0}), which were both present in the No-X-ray run.

In summary, we explain the diverse effects of the X-ray mode on galaxies at different epochs as follows: already before $z=2$ both the No-X-ray and the Simba-50 contain some galaxies with unusually massive and mature black holes for their galaxies. These galaxies likely contained large amounts of gas initially, which resulted in significant black hole growth and high star formation rates. Eventually the accretion rate drops below $f_\mathrm{edd} =0.2$, which activates the jet mode. These are not efficient at quenching star formation by themselves at this point, but eventually gas fractions are low enough for the X-ray mode to turn on, which quenches star formation activity in the central regions. This is seen in the central suppression of the sSFR in Simba-50 galaxies compared to galaxies in the No-X-ray run \citep{Appleby2019TheSimulation}. While the radiative wind and jet modes act in the purely bipolar fashion and do not affect the local ISM significantly, the energy input of the X-ray mode is spherical and directly affects the central gas. All ISM and non-ISM gas particles within the black hole kernel receive a temperature increase, as well as a radial outward momentum kick for just the ISM gas. In addition to dispelling the ISM gas from the central regions, the thermal input may be significant enough to surpass the ISM temperature floor defined in Eq.~\eqref{eq:ISM} and the affected gas is thus no longer eligible for star formation. The resulting quenching of the central star formation is rapid \citep{Montero2019MergersSimulation} and independent of the current accretion rate. Residual star formation outside of the central region will vary among different galaxies, thus explaining the break of the correlation between $f_\mathrm{edd}$ and the overall sSFR's upon the activation of X-ray feedback. Early quenched galaxies with formerly high star formation rates were likely particularly affected due to their large black hole masses, whose increased sphere of influence was able to quench most of the star formation. 

The black hole accretion rate itself is similarly affected by the central thermal energy input. For cold gas, torque-limited accretion is the dominant accretion mode, since it accretes gas below $T < 10^{5} \rm K$, which is primarily in the ISM phase. If gas in the black hole kernel is suddenly heated to higher temperatures, it is accreted via spherical Bondi accretion. This mode is much less stable, resulting in a significantly more variable accretion rate. As a result, the anti-correlation between the Eddington ratio and the black hole mass also breaks down. \citet{Thomas2019BlackSimba} follow a similar reasoning related to the prevalence of hot gas at the expense of cold gas in quenched systems to explain the same feature found in the Simba-100 run. They ascribed the origin of the imperfect anti-correlation between black hole accretion rate and mass to the action of AGN jets. However, as we showed in this manuscript, only upon contrasting different \textsc{Simba} runs isolating the AGN feedback prescriptions, it becomes clear that only the inclusion of the X-ray feedback produces this feature. Similarly, \citet{Szpila2024TheSimulation} examined early ($z\geq2$) quenched galaxies in Simba-100, and by tracking individual systems showed that each early quenching event was preceded by going into jet mode. Our analysis using the \textsc{Simba} variants suggests that it is primarily the influence of the X-ray feedback accompanying AGN jets that drives early quenching.  

\subsection{Comparisons to other cosmological simulations}
Most modern cosmological simulations and galaxy evolution models contain prescriptions for stellar feedback and AGN feedback. Despite some similarities in the modelling approach, each of them differs in the exact parametrisation and underlying assumptions. 
In this section, we compare our results with previous similar attempts at disentangling the effects of different feedback prescriptions on the star formation history predicted by other cosmological simulations.

\citet{Vogelsberger2013APhysics} analyse several variants of a $25 h^{-1}\mathrm{cMpc}$ cosmological box with the galaxy formation model that forms the basis of the \textsc{Illustris} simulations \citep{Vogelsberger2014IntroducingUniverse}. Similar 
to \textsc{Simba}, stellar feedback is implemented kinetically and three distinct prescriptions are used to model AGN feedback. \textsc{Illustris} uses a thermal `quasar mode' at high accretion rates and a kinetic `radio mode' at low accretion rates. Additionally, `electromagnetic' feedback from the excess UV photoionising and photoheating radiation in the proximity of AGN is included in the fiducial run. 
However, since the strength of this mode depends on the luminosity output of the AGN, its effects are only significant in the radiatively efficient quasar mode. \citet{Vogelsberger2013APhysics} then compare the evolution of key galaxy properties across runs with successive inclusions of the feedback modes. Similar to our results from \textsc{Simba}, they find that stellar feedback shapes the cosmic star formation rate density at early times, whereas specifically the low-accretion radio mode takes over at late times. This transition begins around $z=4$, earlier than \textsc{Simba}'s jet mode is able to quench galaxies. However, the inclusion of the X-ray mode in \textsc{Simba} decreases the SFRD to a similar degree after $z=4$, thus the fiducial \textsc{Simba} and \textsc{Illustris} show a similar evolution overall. \citet{Vogelsberger2013APhysics} further find that the GSMF at $z=0$ is shaped by stellar feedback in galaxies below $M_{\star} \sim 10^{11}\mathrm{\, M_{\odot}}$, above which the radio mode takes over. This is again in agreement with the effects of \textsc{Simba}'s stellar and jet modes. The quasar and electromagnetic AGN mode in \textsc{Illustris} have almost no effect on both the SFRD and GSMF. 

\textsc{IllustrisTNG} \citep{Pillepich2018SimulatingModel,Weinberger2017SimulatingFeedback} is the successor to the \textsc{Illustris} simulations. Among other modelling improvements, it contains a more sophisticated stellar feedback model \citep[see][]{Pillepich2018SimulatingModel} and a kinetic radio feedback mode \citep[see][]{Weinberger2017SimulatingFeedback}. The kinetic radio mode generates bipolar jets, but unlike the AGN-jet mode in \textsc{Simba}, their direction is randomised, thus not generally aligned with the rotation axis of the inner galactic disk. The original thermal quasar and the radiative electromagnetic prescriptions remain the same as in \textsc{Illustris}. \citet{Pillepich2018SimulatingModel} compare the galaxy populations in several $25 h^{-1} \mathrm{cMpc}$ runs with isolated mechanisms for stellar and AGN feedback. As in \textsc{Simba} and \textsc{Illustris}, the SFRD evolution is shaped exclusively by stellar feedback before $z=4$, after which AGN feedback becomes relevant and creates a peak at $z=2-3$. \textsc{IllustrisTNG}'s stellar feedback similarly sets the GSMF at $z=0$ below $M_{\star} \sim 10^{11}\mathrm{\, M_{\odot}}$, while AGN feedback regulates the GSMF at higher masses. For the SFRD and GSMF, effects from the quasar and thermal kinetic mode are not considered separately. However, the SMHM ratio shows that quasar mode is at least partially responsible for suppressing stellar mass growth in haloes between $10^{11}$ and $10^{12}\mathrm{\, M_{\odot}}$, whereas the radio mode is needed to create the peak at $10^{12}\mathrm{\, M_{\odot}}$.

The \textsc{Horizon-AGN} \citep{Dubois_2014} simulations models stellar feedback kinetically for the first $50 \, \rm Myr$ and thermally thereafter. As in \textsc{IllustrisTNG}, AGN feedback is once again split into an isotropic thermal `quasar' mode at high-accretion rates and a kinetic jet-producing `radio' mode at low-accretion rates. We note however, that the transition between these modes, as well as the exact implementation differs from that of \textsc{IllustrisTNG}. \citet{Kaviraj2017TheTime} compare the evolution of the GSMF in the fiducial  Horizon-AGN run to the Horizon-NoAGN run, where all AGN feedback is disabled. They find that AGN feedback is crucial to reproduce observed number densities of high-mass galaxies ($M_{\star} > 10^{11}\mathrm{\, M_{\odot}}$) below redshifts $z \sim 2$. 

\textsc{Eagle} \citep{Schaye2015TheEnvironments} is unique among modern cosmological simulations, as both its stellar and its sole AGN feedback mode are modelled as an input of thermal energy. Among a comparison of $50 h^{-1}\mathrm{cMpc}$ \textsc{Eagle} variants based on different cosmologies, \citet{Salcido2018TheHold} show the SFRD evolution in a fiducial and a No-AGN version of the $\Lambda \mathrm{CDM}$ Universe. In the absence of AGN feedback, they find an increased normalisation of around 0.25\,dex after $z=4$ and no change in slope. While the time where AGN feedback becomes relevant is in agreement with other simulations, this is a much weaker effect than in \textsc{Simba}, \textsc{Illustris}, and \textsc{IllustrisTNG}. In those simulations, the continued quenching due to AGN feedback produces a suppression of at least 0.5\,dex at $z=0$ and is needed to move the peak of the SFRD between $z=2$ and 3, as well as steepen its late-time slope. \citet{Salcido2018TheHold} only show the GSMF at $z=0.1$, but they find the same mass cut-off at $M_{\star} = 10^{11}\mathrm{\, M_{\odot}}$, above which AGN feedback suppresses the number density of galaxies.

Overall, the cosmological simulations \textsc{Illustris}, \textsc{IllustrisTNG}, \textsc{Horizon-AGN}, \textsc{Eagle} and \textsc{Simba} largely find that their stellar and AGN feedback modes affect global galaxy statistics in a similar way, despite differing implementations and underlying assumptions. 
This converges to a unified picture where the SFRD is shaped by stellar feedback at early times, while AGN feedback starts showing effects after $z=4$ and becomes the dominant quenching mechanism after $z=2$. AGN feedback produces a sharp turnover at $M_{\star} = 10^{11}\mathrm{\, M_{\odot}}$ in the GSMF, while below this threshold stellar feedback suppresses the number density of galaxies. A feature that is introduced with kinetic AGN feedback modes is a steeper decrease of the SFRD at late times. However, it does not appear to be crucial in order to reproduce the observed SFRD. Indeed, \textsc{Eagle} yields a good match with the observed SFRD by using a single, purely radiative, AGN feedback prescription. 

The overall similarity of the conclusions that can be drawn from different models suggests that the emerging picture of star formation history in a cosmological context is robust. However, this makes it harder to discriminate amongst different feedback prescriptions with observations of global properties such as the SFRD or GSMF. A more promising strategy would be focusing on potentially more sensitive observables. A good example is the evolution of the $f_{\rm edd} - M_{\rm BH}$ relationship, which we showed to be highly sensitive to X-ray feedback. A complementary approach would be to track the evolution of the star formation history of individual galaxies in key mass ranges, and especially across $z=2$, which acts as a watershed between a stellar-feedback and AGN-feedback-dominated eras for high-mass haloes. Indeed, even if different models broadly agree on the global star formation history, they may still differ in their predictions for restricted populations of galaxies. We will address these questions in future work.

\section{Conclusions}
\label{sec:conclusions}
In this work, we investigated the impact of stellar and AGN feedback on the cosmic star formation history using the cosmological \textsc{Simba} simulation suite. By comparing the redshift evolution of key observables across different runs of the 50 $h^{-1} \mathrm{cMpc}$ run with isolated feedback variants, we have successfully connected the overall quenching seen in global statistics to feedback mechanisms acting in individual haloes and galaxies. This enabled us to form a cohesive picture of how \textsc{Simba}'s feedback modes have shaped the cosmic star formation history:

\begin{itemize}
    \item While AGN feedback starts quenching star formation in individual haloes after $z=4$, $z=2$ emerges as a key redshift transition, after which AGN feedback modes take over as the dominant quenching mechanisms. Before $z=2$, the shape of global statistics like the SFRD (Fig. \ref{fig:SFRD}) and the GSMF (Fig. \ref{fig:GSMF}) is set by stellar feedback. This conclusion is analogous to the findings by \cite{Sorini2022} on the impact of feedback modes on the distribution of gas within haloes, in the CGM, and in the IGM, hence underscoring the tight correlation between the effect of feedback on the gaseous and stellar components.
    
    \item The extent to which different haloes are affected by each feedback mode strongly depends on their mass.  
    Stellar feedback shapes both the evolution of the instantaneous SFR (see Fig. \ref{fig:halobins}), as well as the associated stellar mass growth (Fig. \ref{fig:heatmapz2} and \ref{fig:heatmapz0}), in low-mass haloes of $\lesssim 10^{12} \rm \, M_{\odot}$, which are also more common at $z>2$. In these haloes, the gravitational potential is too low to compete with the speeds of mass-loaded stellar winds, leading to outflows large enough to quench the star formation rate. In larger haloes, above $10^{12} \rm \, M_{\odot}$, stellar feedback becomes less effective due to an increased gravitational potential and a decreased mass-loading factor in \textsc{Simba}. These haloes tend to contain massive black holes, which have matured to low accretion rates, thus enabling efficient quenching due to AGN feedback.
     
    \item Among the three AGN feedback modes in \textsc{Simba}, jet feedback at low accretion rates is by far the most effective at quenching star formation. This is largely a slow quenching mechanism, where the jets heat the surrounding CGM to high temperatures, which prevents further gas accretion. 
    The radiative AGN winds alone at high accretion rates are only relevant for quenching intermediate-mass galaxies between $M_{\star} \sim 10^{9.5}$ and $10^{10}\mathrm{\, M_{\odot}}$. X-ray heating mainly quenches residual star formation in the central regions, which at late times mainly affects the number of massive galaxies between $10^{11}$ and $10^{12}\mathrm{\, M_{\odot}}$, as well as their SFRs, which have already been quenched by jets.

	\item We further found new that the X-ray heating mode, in particular, is necessary to produce the first fully quenched massive galaxies before $z=2$, where jets alone are not yet efficient (Fig. \ref{fig:SFR_Mstar 2} and \ref{fig:SFR_Mstar 0}). As soon as galaxies become eligible for X-ray feedback, the gas temperature around the black hole is increased and/or the gas is pushed outwards spherically. For a few galaxies with a particularly massive black hole and a centrally concentrated star formation, this produces the earliest galaxies with quenched sSFR's below $10^{-10}\mathrm{yr^{-1}}$. As both black hole growth and star formation draw from the same gas reservoir, this small number of galaxies initially showed the highest star formation rates prior to quenching. Their quenching is significant enough to manifest in a suppression of the overall SFRD already after $z=4$ (Fig. \ref{fig:SFRDSFR}). 
    \item The X-ray mode is also responsible for breaking the anti-correlation between the Eddington ratio and the black hole mass (Fig. \ref{fig:fedd_MBH}), which had been identified in the fiducial Simba-100 run by \citep{Thomas2019BlackSimba}. 
    The heating of the central gas likely leads to a preferential accretion via the spherical Bondi mode, which is much less stable than the torque-limited accretion from cold gas. As a result, the accretion rates are more variable for any given black hole mass and no longer correlated with the specific SFR's.
\end{itemize}

Comparing our results with previous similar works utilising other cosmological simulations \citep[e.g.][]{Vogelsberger2013APhysics, Vogelsberger2014IntroducingUniverse, Weinberger2017SimulatingFeedback, Pillepich2018SimulatingModel, Salcido2018TheHold}, we find broad agreement on the effect of different feedback modes on global observables of the star formation history, such as the SFRD and GSMF. However, the inclusion of a kinetic AGN feedback mechanism at low-accretion rates in simulations that adopt multiple feedback modes appears to be necessary in order to match the late-time slope of the observed SFRD. The \textsc{Eagle} simulation represents an exception, as it reproduces the data with a single radiative AGN feedback mode only.

Despite the agreement on global observables, different models may still predict different evolutions of the properties of individual galaxies. For instance, the aforementioned models produce significantly different cool gas contents within galaxies~\citep{Dave_2020} and predictions for sub-millimetre galaxies~\citep{Lovell2021ReproducingSimulations}.
We wish to explore more such discriminating scenarios in future work. We further want to assess the feasibility of observationally confirming or rejecting the characteristic predictions of \textsc{Simba} for the effect of X-ray feedback in the accretion-rate-mass relationship of supermassive black holes.

\section*{Acknowledgements}

We are grateful to Nicole Thomas and the members of the \textsc{Simba} collaboration for helpful discussions. We acknowledge the \texttt{yt} team for development and support of \texttt{yt}. DS was supported by the by the STFC consolidated grant no. RA5496. 
This work used the DiRAC\MVAt Durham facility managed by the Institute for Computational Cosmology on behalf of the STFC DiRAC HPC Facility. The equipment was funded by BEIS capital funding via STFC capital grants ST/P002293/1, ST/R002371/1 and ST/S002502/1, Durham University and STFC operations grant ST/R000832/1. DiRAC is part of the National e-Infrastructure. 
This work made extensive use of the NASA Astrophysics Data System and of the astro-ph preprint archive at arXiv.org. For the purpose of open access, the author has applied a Creative Commons Attribution (CC BY) licence to any Author Accepted Manuscript version arising from this submission.

%%%%%%%%%%%%%%%%%%%%%%%%%%%%%%%%%%%%%%%%%%%%%%%%%%
\section*{Data Availability}

The simulation data underlying this article are publicly available\footnote{\url{http://simba.roe.ac.uk}}. 
The derived data will be shared upon reasonable request to the corresponding author. 

%The inclusion of a Data Availability Statement is a requirement for articles published in MNRAS. Data Availability Statements provide a standardised format for readers to understand the availability of data underlying the research results described in the article. The statement may refer to original data generated in the course of the study or to third-party data analysed in the article. The statement should describe and provide means of access, where possible, by linking to the data or providing the required accession numbers for the relevant databases or DOIs.

%%%%%%%%%%%%%%%%%%%% REFERENCES %%%%%%%%%%%%%%%%%%

% The best way to enter references is to use BibTeX:

\bibliographystyle{mnras}
\bibliography{bib.bib} % if your bibtex file is called example.bib

% Alternatively you could enter them by hand, like this:
% This method is tedious and prone to error if you have lots of references
%\begin{thebibliography}{99}
%\bibitem[\protect\citeauthoryear{Author}{2012}]{Author2012}
%Author A.~N., 2013, Journal of Improbable Astronomy, 1, 1
%\bibitem[\protect\citeauthoryear{Others}{2013}]{Others2013}
%Others S., 2012, Journal of Interesting Stuff, 17, 198
%\end{thebibliography}

%%%%%%%%%%%%%%%%%%%%%%%%%%%%%%%%%%%%%%%%%%%%%%%%%%

%%%%%%%%%%%%%%%%% APPENDICES %%%%%%%%%%%%%%%%%%%%%

\appendix

\section{Numerical convergence}
\label{app:A}

\begin{figure}
    \centering
    \includegraphics[width=\columnwidth]{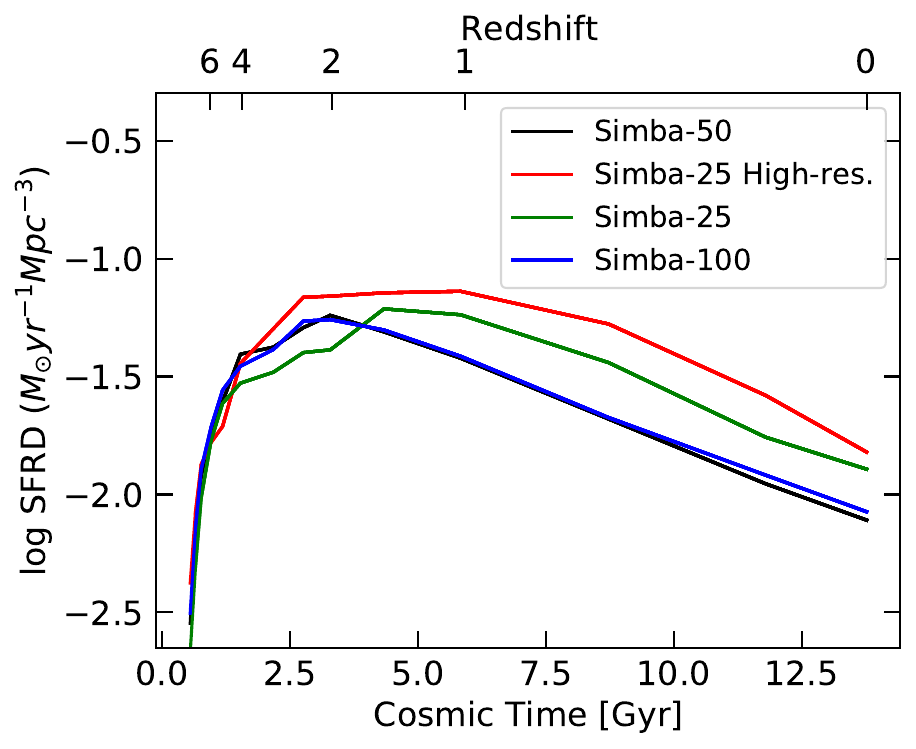}
    \caption{Convergence test for the SFRD. We compare the Simba-50 fiducial run (black) to the Simba-100 (blue), Simba-25 (green) and the Simba-25 High res run (red). While volume-wise convergence between Simba-50 and Simba-100 is excellent, the Simba-25 versions produce slightly increased SFRD's.}
    \label{appendix:conv} 
\end{figure}

Figure \ref{appendix:conv} shows a volume and resolution convergence test for the SFRD using the different simulations with full physics \textsc{Simba} models. We contrast Simba-100 (blue), Simba-50 (black), Simba-25 (green), and Simba-25 High-res runs (red). We refer to Table \ref{tab:runs} for the specifications of each run.

Comparing runs with the same resolution, we note excellent volume-wise convergence between the Simba-100 model and Simba-50. Simba-25 exhibits a higher SFRD by around 0.2~dex, which is likely because the small volume produces fewer massive haloes. As a result, the AGN jet feedback is not well-sampled, which produces the most significant suppression in the SFRD. The resolution-wise convergence between the Simba-25 and the Simba-25 High-res run is not as close. The SFRD in the High-res run is increased by 0.1--0.25\,dex with respect to the Simba-25 run, as small haloes, in particular, are better resolved.

We note, however, that the purpose of this work is to isolate the relative strength of \textsc{Simba}'s feedback modes, rather than qualifying results from the fiducial model. Since the runs with different feedback variants are only available for Simba-50, we unfortunately cannot test the volume- and resolution dependence of the effects due to each feedback mode.

\
%%%%%%%%%%%%%%%%%%%%%%%%%%%%%%%%%%%%%%%%%%%%%%%%%%

% Don't change these lines
\bsp	% typesetting comment
\label{lastpage}
\end{document}